\documentclass[longauth]{aa} % for the long lists of affiliations 
\usepackage{graphicx}
\usepackage{hyperref}
\usepackage{txfonts}
\usepackage{siunitx}
\begin{document}

   \title{FAUST XXV. A potential new molecular outflow in [BHB2007]\,11}

   \subtitle{}

   \author{A. Martínez-Henares \inst{1,2}
          \and
          I. Jiménez-Serra \inst{1}
          \and
          C. Vastel \inst{3}
          \and
          T. Sakai \inst{4}
          \and 
          L. Evans \inst{5}
          \and
          J. E. Pineda \inst{6}
          \and
          M. J. Maureira \inst{6}
          \and
          E. Bianchi \inst{7}
          \and
          C. J. Chandler \inst{8}
          \and
          C. Codella \inst{7}
          \and
          M. De Simone \inst{9}
          \and 
          L. Podio \inst{7}
          \and
          G. Sabatini \inst{7}
          \and
          Y. Aikawa \inst{10}
          \and
          F. O. Alves \inst{11}
          \and
          M. Bouvier \inst{12}
          \and
          P. Caselli \inst{6}
          \and
          C. Ceccarelli \inst{13}
          \and
          N. Cuello \inst{13}
          \and
          F. Fontani \inst{7}
          \and
          T. Hanawa \inst{14}
          \and 
          D. Johnstone \inst{15,16}
          \and
          L. Loinard \inst{17,18,19}
          \and
          G. Moellenbrock \inst{8}
          \and
          S. Ohashi \inst{20}
          \and
          N. Sakai \inst{21}
          \and
          D. Segura-Cox \inst{6}
          \and
          B. Svoboda \inst{8}
          \and
          S. Yamamoto \inst{22}
          }

   \institute{Centro de Astrobiología (CAB), INTA-CSIC, Carretera de Ajalvir km 4, Torrejón de Ardoz, 28850, Madrid, Spain %1
               \and 
                    Escuela de Doctorado, Universidad Autónoma de Madrid, 28049 Cantoblanco, Madrid, Spain %2
                \and 
                    IRAP, Université de Toulouse, CNRS, CNES, UPS, Toulouse, France %3
                \and
                    Graduate School of Informatics and Engineering, The University of Electro-Communications, Chofu, Tokyo 182-8585, Japan %4
                \and
                    School of Physics and Astronomy, University of Leeds, Leeds LS2 9JT, UK %5
                \and
                    Center for Astrochemical Studies, Max-Planck-Institut für Extraterrestrische Physik, Gießenbachstraße 1, 85748 Garching, Germany %6
                \and
                    INAF – Osservatorio Astrofisico di Arcetri, Largo E. Fermi 5, 50125, Florence, Italy %7
                \and
                    National Radio Astronomy Observatory, PO Box O, Socorro, NM 87801, USA %8
                \and
                    European Southern Observatory, Karl-Schwarzschild-Strasse 2 D-85748 Garching bei Munchen, Germany %9
                \and
                    Department of Astronomy, The University of Tokyo, 7-3-1 Hongo, Bunkyoku, Tokyo 113-0033, Japan %10
                \and
                    IRAM, 300 rue de la Piscine, F-38406 Saint-Martin d’Hères, France %11
                \and
                    Leiden Observatory, Leiden University, PO Box 9513, 23000 RA Leiden, The Netherlands %12
                \and
                    Univ. Grenoble Alpes, CNRS, IPAG, 38000 Grenoble, France %13
                \and
                    Center for Frontier Science, Chiba University, 1-33 Yayoi-cho, Inage-ku, Chiba 263-8522, Japan %14
                \and
                    NRC Herzberg Astronomy and Astrophysics, 5071 West Saanich Road, Victoria, BC, V9E 2E7, Canada %15
                \and
                    Department of Physics and Astronomy, University of Victoria, Victoria, BC, V8P 5C2, Canada %16
                \and
                    Instituto de Radioastronomía y Astrofísica, Universidad Nacional Autonóma de México, Apartado Postal 3-72, Morelia 58090, Michoacán, Mexico %17
                \and
                    Black Hole Initiative at Harvard University, 20 Garden Street, Cambridge, MA 02138, USA %18
                \and
                    David Rockefeller Center for Latin American Studies, Harvard University, 1730 Cambridge Street, Cambridge, MA 02138, USA %19
                \and
                    National Astronomical Observatory of Japan, Osawa 2-21-1, Mitaka-shi, Tokyo 181-8588, Japan %20
                \and
                    RIKEN Cluster for Pioneering Research, 2-1, Hirosawa, Wako-shi, Saitama 351-0198, Japan %21
                \and
                    SOKENDAI, Shonan Village, Hayama, Kanagawa 240-0193, Japan %22
              }
         % \and
         %     University of Alexandria, Department of Geography, ...\\
         %     \email{c.ptolemy@hipparch.uheaven.space}
         %     \thanks{The university of heaven temporarily does not
         %             accept e-mails}
         %     }

   \date{}

% \abstract{}{}{}{}{} 
% 5 {} token are mandatory
 
  \abstract
  % context heading (optional)
  % {} leave it empty if necessary  
   {During the early stages of star formation, accretion processes such as infall from the envelope and molecular streamers, and ejection of matter through winds and jets, take place simultaneously and distribute the angular momentum of the parent molecular cloud. The Class 0/I binary [BHB2007]\,11 shows evidence for accretion and ejection at the scales of the circumbinary disk and the inner close binary. Recent observations of H$_2$CO, however, show two elongated structures with hints of outflowing motion almost perpendicular to the main CO outflow, which is launched from the circumbinary disk.}
  % aims heading (mandatory)
   {Our aim is to study the kinematics of the molecular gas at intermediate scales of $\sim50-3000$\,au around [BHB2007]\,11 to verify the nature of these elongated structures.}
  % methods heading (mandatory)
   {We analyze the line emission of H$^{13}$CO$^+$, CCH, c-C$_3$H$_2$ and SiO observed with the Atacama Large Millimeter/submillimeter Array (ALMA) within the Large Program FAUST (Fifty AU STudy of the chemistry in the disc/envelope system of Solar-like protostars). These molecules trace the material moving at velocities close to the ambient cloud velocity, which could not be probed in previous observations of the self-absorbed emission of CO.}
  % results heading (mandatory)
   {The images of H$^{13}$CO$^+$, CCH, c-C$_3$H$_2$ show clear elongated structures similar to those previously detected in H$_2$CO, whose gas kinematics are consistent with outflowing motions and with rotation in the opposite sense to the main CO outflow. The derived mass loss rate from these large-scale structures is $(1.8\pm0.5)\times10^{-6}M_{\odot}\textrm{ yr}^{-1}$, in agreement with those measured in outflows driven by Class 0/I protostars. The SiO image reveals compact emission close to the binary system, with a slight elongation aligned with the larger-scale structures. This suggests that SiO is released from the sputtering of dust grains in the shocked material at the base of the potential new outflow, with a relative abundance of $\geq(0.11-2.0)\times10^{-9}$. However, higher angular and spectral resolution observations are needed to accurately estimate the outflow launching radius and its powering source. Given the location and the abundance of the SiO emission, we propose that the second outflow may be launched from inside the circumbinary disk, likely by the less massive companion, which is actively accreting material from its surroundings.}
  % conclusions heading (optional), leave it empty if necessary 
   {}


   \maketitle
%
%-------------------------------------------------------------------

\section{Introduction}

Star formation starts with the gravitational collapse of a molecular cloud \citep{shu1977}. Due to angular momentum conservation, the infalling envelope of material eventually flattens and gives rise to a circumstellar disk of gas and dust \citep{terebey1984}. The disk rotates around the central protostar, whose mass grows as it accretes material from the circumstellar disk. These accretion processes are facilitated by the removal of angular momentum from the disk through the launching of winds and jets \citep[e.g.][]{frank2014}. In recent years, there has been observational effort focused on determining the mechanism that regulates these ejection processes by studying their kinematics \citep[e.g.][]{rayferreira2021}. At the same time, accretion onto the protostar has been observed to happen through the infall of the rotating, flattened envelope that provides material to the accretion disk through a centrifugal barrier \citep[e.g.][]{sakai2014,oya2016} and, more recently, through dense filamentary streamers that funnel material from the surroundings of the protostar \citep[see e.g. the recent review by][]{pineda2023}. To date, it is still not clear how these accretion and ejection processes interplay and how the angular momentum of the system is distributed during the early stages of star formation.

Most stars are found in binary systems. During their early formation and evolution, each member of the binary hosts its own circumstellar disk and outflow \citep[e.g.][]{offner2023}. Circumbinary disks surrounding the central sources may be present as well: they are observed to be more frequent for binary separations less than $\sim$10-20 au \citep{czekala2019}. Some notable examples are found at the Class 0/I stage of a protostar, such as VLA1623AB \citep[e.g.][]{hara2021,ohashi2022,codella2024}, L1551 IRS 5 \citep[e.g.][]{rodriguez2003,bianchi2020}, IRAS 16293-2422 A \citep[e.g.][]{maureira2020} and [BHB2007]\,11 \citep{alves2017,alves2019}. These complex dynamical systems are ideal targets to study the kinematics of the gas components involved during the formation of typical stars, and yield a powerful tool to address open questions such as the wind and jet launching mechanism or the relative alignment between circumstellar and circumbinary disks.

In particular, [BHB2007]\,11 represents an excellent laboratory because it has the first evidence for the presence of outflows driven by a circumbinary disk \citep{alves2017}. Also known as B59\#11, it is a Class 0/I protostellar object located at a distance of 163 $\pm$ 5 pc \citep{dzib2018}. With an age of 0.1-0.2\,Myr and a bolometric luminosity of 2.2$-$4.5\,$L_{\odot}$ \citep{brooke2007,forbrich2009,sandell2021}, it is the youngest member of a cluster of at least 20 low-mass young stellar objects in the Barnard 59 (B59) core. Observations with the Atacama Large Millimeter/submillimeter Array (ALMA) at 1.3 mm with a spatial resolution of $\sim$36\,au revealed a circumbinary disk embedded within a flattened envelope that spans \mbox{$\sim$435 au}, where both structures were found to rotate in Keplerian motion \citep{alves2017}. The \mbox{CO\,(2-1)} emission revealed an outflow in the NE-SW direction that was observed to originate from the circumbinary disk at a launching radius of \mbox{$\sim$90-130 au} from the central system. This radius coincides with the centrifugal barrier and with the tip of spiral structures seen in thermal dust emission associated with the accretion from the surrounding flattened envelope. The infall of matter at the centrifugal barrier drags the magnetic field lines and pinches their opening angle, allowing the magneto-centrifugal launching of the outflow from the circumbinary disk \citep{alves2017}. \cite{alves2018} presented Very Large Array (VLA) data at 34.5\,GHz obtained in its most extended configuration that resolved the two members of the system in the inner part of the circumbinary disk, [BHB2007]\,11A ($\alpha$(2000) = 17h11m23.1058s, $\delta$(2000) = $-$27$^{\circ}$24'32.828") and [BHB2007]\,11B ($\alpha$(2000) = 17h11m23.1015s, $\delta$(2000) = $-$27$^{\circ}$24'32.987"). Subsequent ALMA observations at \mbox{1.3 mm} with an angular resolution of $\sim$6\,au \citep{alves2019} imaged the circumstellar disks of the two members of the system with a projected separation of $\sim$28\,au, and unveiled filamentary spiral features with an extent of \mbox{$\sim$100 au} that connected the close binary to the circumbinary disk. In a reanalysis of the \mbox{CO\,(2-1)} data from \cite{alves2017}, the authors obtained the centroid positions of the high velocity wings. The line emission was found to be situated at scales of the inner circumbinary disk, in the surroundings of source B, with an acceleration motion towards the protostar. This was interpreted as evidence of accreting material from the circumbinary disk to the two circumstellar disks through the spiral filaments, which act as streamers.

[BHB2007]\,11 has recently been re-imaged with ALMA as part of the sample studied within the ALMA Large Program FAUST \citep[Fifty AU STudy of the chemistry in the disc/envelope system of Solar-like protostars,][]{codella2021}. These observations have already provided additional insight into this system: \cite{vastel2022} and \cite{vastel2024} have reported compact emission from complex organic molecules (such as methanol, methyl formate and dimethyl ether) in the inner circumbinary disk and surrounding source B, which may be related to shocked gas from the filamentary structures surrounding both cores. Still, further observations would be needed to address the effect of very optically thick dust emission from the two sources \citep[see][]{desimone2020} and confirm the origin of the emission from accretion shocks. \cite{evans2023} studied the deuteration ratio of H$_{2}$CO in this system and found that the deuterated species are located in the central regions of the core. This work also reported the presence of two elongated H$_{2}$CO structures detected at large scales outside the circumbinary disk in the SE-NW direction. These structures were detected at velocities of $\pm$1 km~s$^{-1}$ from the systemic velocity of the cloud $v_{sys}\sim$3.6 km~s$^{-1}$ \citep{hara2013}, and they are aligned almost perpendicular to the main CO outflow reported by \cite{alves2017} using the CO\,(2-1) transition. The kinematic analysis of these gas structures suggested that they were not infalling streamers since they could not be fitted by streamline-generating models \citep{evans2023}, which predict increasing velocities towards the central protostar \citep{pineda2020}. Instead, the kinematics of the blueshifted H$_{2}$CO emission are consistent with gas being accelerated radially outwards, mimicking the kinematics of an outflow. The redshifted emission presented a flat velocity structure that complicated the interpretation of the origin of the H$_{2}$CO emission. 

In this work, we analyze FAUST observations of the large-scale molecular emission from [BHB2007]\,11 in search for more information on the nature of the recently discovered H$_{2}$CO elongated structures which may be associated with a potential second outflow in the [BHB2007]\,11 system. We also revisit ALMA archival observations of CO\,(3-2) and compare them with the FAUST data. Our analysis of the FAUST images of H$^{13}$CO$^+$, CCH and c-C$_3$H$_2$ reveals the elongated structures and the outflowing motions, in agreement with the outflow origin of this emission. The FAUST images also detect compact, faint and slightly elongated SiO emission associated with the inner binary system, whose alignment coincides with that observed for H$^{13}$CO$^+$, CCH, c-C$_3$H$_2$, and that likely arises from shocked material at the base of the second outflow. 

%--------------------------------------------------------------------
\section{Observations and data reduction}

[BHB2007]\,11 was observed with ALMA under the FAUST Large Program (project code 2018.1.01205.L) between 2018 and 2020. Observations, data reduction and imaging are described in detail in previous FAUST articles on the source \citep{vastel2022,evans2023,vastel2024}. The observations were centered at $\alpha$(2000)=17h11m23.125s, $\delta$(2000)=–27$^{\circ}$24'32.87''. The spectral setup consisted of two frequency setups (setups 1 and 2) in Band 6 and one (setup 3) in Band 3. The molecular line transitions studied in this work were covered in setup 1 and 2 with a spectral channel width of 122.07 kHz (equivalent to a velocity resolution of 0.14 km~s$^{-1}$). Observations for the two relevant setups were performed with the 12~m and 7~m antennas, with baseline lengths from 15~m to 1.3~km for the 12m array and from 8.9~m to 48.9~m for the 7~m array. The calibration and imaging of the data was performed with CASA using a modified version of the pipeline developed by the FAUST data reduction team \footnote{\url{https://faust-imaging.readthedocs.io/en/latest/}}. The resulting beam size after applying Briggs weighting with a robust parameter of 0.5 is $\sim0.42$" across the two setups. The images have been corrected from the primary beam response. In addition to the line cubes obtained under the FAUST program, we have retrieved the CO\,(3-2) line cube from the ALMA archive (project 2019.1.01566.S). The imaging of these visibilities has been done by the ALMA pipeline with a Briggs parameter of 0.5 and primary beam correction. The cube has a spectral resolution of 0.24~MHz (equivalent to $\sim$0.11 km~s$^{-1}$) and a synthesized beam size of 0.11".
%--------------------------------------------------------------------

\section{Results}\label{sec:results}

    \subsection{Morphology of the emission}\label{sec:morphology}

    \subsubsection{Large scale structures} \label{sec:elongations}

    \begin{figure*}[ht!]
       \centering
       \includegraphics[width=0.865\textwidth]{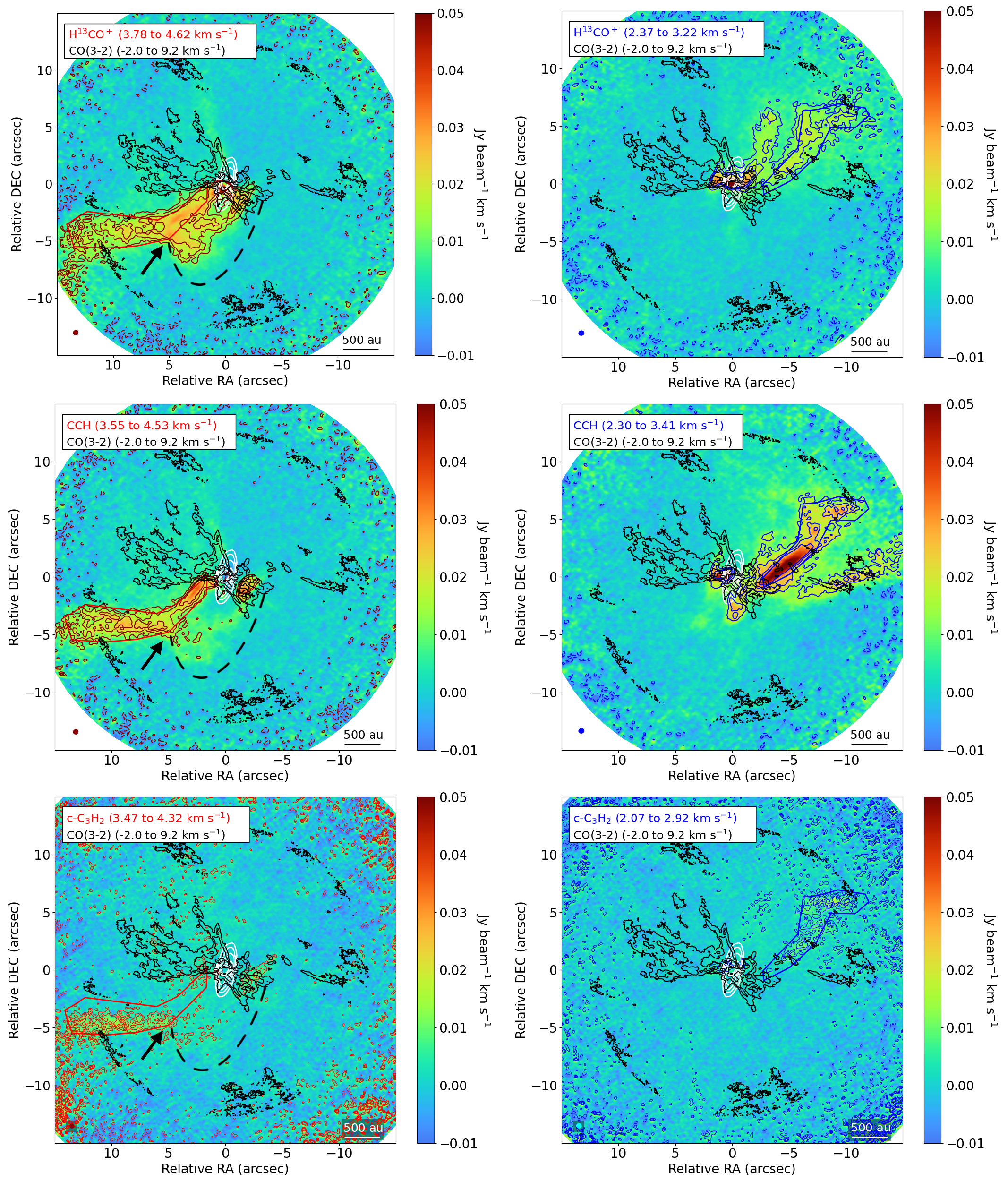}
       \caption{Integrated intensity maps of H$^{13}$CO$^+$ (top panels), CCH (middle) and c-C$_3$H$_2$ (bottom) in colorscale and colored contours. Left panels correspond to redshifted velocities from the systemic velocity, $v_{sys}\sim$3.6 km~s$^{-1}$, and right panels to blueshifted velocities: ranges are specified in the text insets of each panel. Overlaid on all panels we plot the 1.3 mm ALMA continuum (white contours) and the CO\,(3-2) integrated emission (black contours) from $-$2.0 to 9.2 km~s$^{-1}$. H$^{13}$CO$^+$ and CCH contours are \mbox{[-5,10,15,20]} times the 1$\sigma$ rms noise level, which is 1.0\,mJy~beam$^{-1}$~km~s$^{-1}$ for both maps of H$^{13}$CO$^+$, and 1.1\,mJy~beam$^{-1}$~km~s$^{-1}$ for both CCH maps. c-C$_3$H$_2$ contours are \mbox{[-5,3,5,7,10]} times the 1$\sigma$ rms noise level, which is 1.8\,mJy~beam$^{-1}$~km~s$^{-1}$ for both maps. Negative contours are represented with dashed lines. CO\,(3-2) contours are [50,100,200] times the 1$\sigma$ rms noise of 1.2\,mJy~beam$^{-1}$~km~s$^{-1}$. The rms noise of the integrated intensity maps has been measured before correcting for the primary beam response. 1.3 mm continuum contours are [20,30,40,50,100,200] times the 1$\sigma$ rms noise of 0.14 mJy~beam$^{-1}$. The dashed curve and arrows in redshifted panels indicate the transition region between the elongated structure (east, outside from the dashed curve) and the rotating envelope (west, delineated by the dashed curve). The red and blue polygons cover the area used to obtain the PV diagrams shown in Figure \ref{fig:pvsoutflow}. The synthesized beam sizes are 0.34"$\times$0.33" for the continuum emission, 0.13"$\times$0.10" for CO\,(3-2), 0.46"$\times$0.39" for H$^{13}$CO$^+$ and CCH, and 0.42"$\times$0.39" for c-C$_3$H$_2$. The beam ellipses for the H$^{13}$CO$^+$, CCH and c-C$_3$H$_2$ maps are plotted in the bottom left corner of the corresponding panels. The center of the images is at $\alpha$(2000)=17h11m23.125s, $\delta$(2000)=–27$^{\circ}$24'32.87".
       }\label{fig:moment0}
    \end{figure*}

    Figure \ref{fig:moment0} shows the integrated intensity (moment 0) maps of H$^{13}$CO$^+$ (J=3-2, F$_1$=5/2-3/2, F=2-1, $E_u=24.98$\,K) at 260.255 GHz (upper panels), CCH (N=3-2, J=7/2-5/2, F=4-3, $E_u=25.15$\,K) at 262.004 GHz (middle panels) and \mbox{c-C$_3$H$_2$} (6$_{(0, 6)}$-5$_{(1, 5)}$, $E_u=38.61$\,K) at 217.822 GHz (bottom panels) for the redshifted (left panels) and blueshifted (right) velocities with respect to the systemic velocity of the source ($v_{sys}\sim3.6$ km~s$^{-1}$). The velocity ranges in the figure have been selected after inspecting the data cubes channel per channel to obtain the clearest view of the structures. Overlaid on the molecular emission, we show the \mbox{1.3 mm} continuum emission \citep[white contours,][]{alves2017} and moment 0 maps of CO\,(3-2) (black contours) . 
    
    The continuum traces the inner envelope of radius $\sim435$ au with a position angle (PA) of 167$^{\circ}$ \citep{alves2017}. The contours of the CO\,(3-2) integrated emission between -2.0 and 9.2 km~s$^{-1}$ show the biconical NE-SW outflow identified in CO\,(2-1) by \cite{alves2017}. To obtain the PA of the CO\,(3-2) outflow axis, we have performed a linear fit of the outermost contours that delineate the outflow in Fig. \ref{fig:moment0}. The NE lobe has a PA of $58\pm1^{\circ}$, and the SW lobe has a PA of $238\pm2^{\circ}$. Hence, we derive a CO outflow axis of $58\pm2^{\circ}$, almost perpendicular to the orientation of the continuum. The SW lobe of the outflow is redshifted between 5.4 and 12.5 km~s$^{-1}$ before showing significant absorption, and also shows blueshifted emission between -4.0 and 1.6 km~s$^{-1}$ over the same region. The NE lobe shows redshifted emission in its southern part between 5.4 and 16 km~s$^{-1}$, and blueshifted emission in the northern half between -10 and 1.6 km~s$^{-1}$. The morphology of the CO\,(3-2) emission is similar to that of the CO\,(2-1) presented by \cite{alves2017}, who associated the redshifted emission of the NE lobe with a weaker component of the outflow. The fact that both lobes of the CO outflow show blueshifted and redshifted emission suggests that its orientation is close to the plane of the sky \citep{alves2017}. The north-south velocity gradient of the NE lobe was interpreted as outflow rotation from the CO\,(2-1) observations of \cite{hara2013} with the Submillimeter Array (SMA), while the SW lobe showed only redshifted emission in their images. At velocities $\pm$2 km~s$^{-1}$ from the systemic velocity of $\sim3.6$ km~s$^{-1}$, the CO\,(3-2) emission is largely self-absorbed, which prevents studying the morphology of the gas with this line at low velocities. This is the case as well for the CO\,(2-1) emission from \cite{alves2017}.

    In contrast, H$^{13}$CO$^+$, CCH and c-C$_3$H$_2$ present marginal or no self-absorption at velocities close to the systemic velocity of the source. This enables us to explore the morphology and kinematics of the gas in a velocity range that has not been explored before with CO. The integrated emission of H$^{13}$CO$^+$ shows two distinct components, with different kinematics and spatial distributions. The inner $\sim$10\,", marked in the left panels of Fig. \ref{fig:moment0} with the dashed ellipse, correspond to the emission from the extended rotating envelope. The PA of this structure is aligned with the PA of the continuum, and shows the rotation pattern of the envelope: redshifted in the south and blueshifted in the north \citep{alves2017}. At larger distances, however, we find two elongated structures toward the SE and the NW, similar to the ones identified by \cite{evans2023} in H$_2$CO. We indicate with an arrow in the left panels of Fig. \ref{fig:moment0} the transition region between the envelope and the elongated structure. The redshifted CCH emission at the inner $\sim$10\," also shows the rotation of the envelope. The blueshifted emission appears in the northern part of the rotating envelope, while we also find some blueshifted emission in the south. This southern emission arises from the more redshifted part of the adjacent CCH transition (N=3-2, J=7/2-5/2, F=3-2) at 262.006 GHz due to its hyperfine structure, which is separated $\sim$2.2 MHz (i.e. $\sim$2.5 km~s$^{-1}$) from the line that we are using. The two lines are slightly blended at their highest velocities, but the residual emission from the F=3-2 transition does not affect our analysis of the F=4-3 transition at velocities very close to the systemic velocity. We also find emission from this molecule in the base of the SW lobe of the CO\,(3-2) outflow at redshifted velocities, as well as for the blueshifted emission in the NE lobe. Therefore, CCH may be tracing also the cavity walls from the primary CO outflow. Lastly, the lower panels show the integrated emission from c-C$_3$H$_2$, which can trace regions illuminated by UV radiation, such as photo-dissociation regions \citep[PDRs;][]{pety2005}, the envelope of warm carbon chain chemistry sources \citep{sakai2013}, and outflow cavity walls \citep[e.g.][]{murillo2018,tychoniec2021}. The c-C$_3$H$_2$ emission is mainly seen in the elongated structures, beyond the transition region where the rotating envelope is seen in the other molecular tracers. As in the case of CCH,  faint emission is found at the base of the CO\,(3-2) outflow, suggesting that this molecule is also tracing its cavity walls. Based on the morphology of the integrated emission of the three tracers, we define two regions (colored polygons in Fig. \ref{fig:moment0}) that we employ for the kinematic analysis of the gas in Sect. \ref{sec:kinematics}.

    \subsubsection{Compact SiO emission} \label{sec:siomorphology}

    \begin{figure*}[h]
       \centering
       \includegraphics[width=\textwidth]{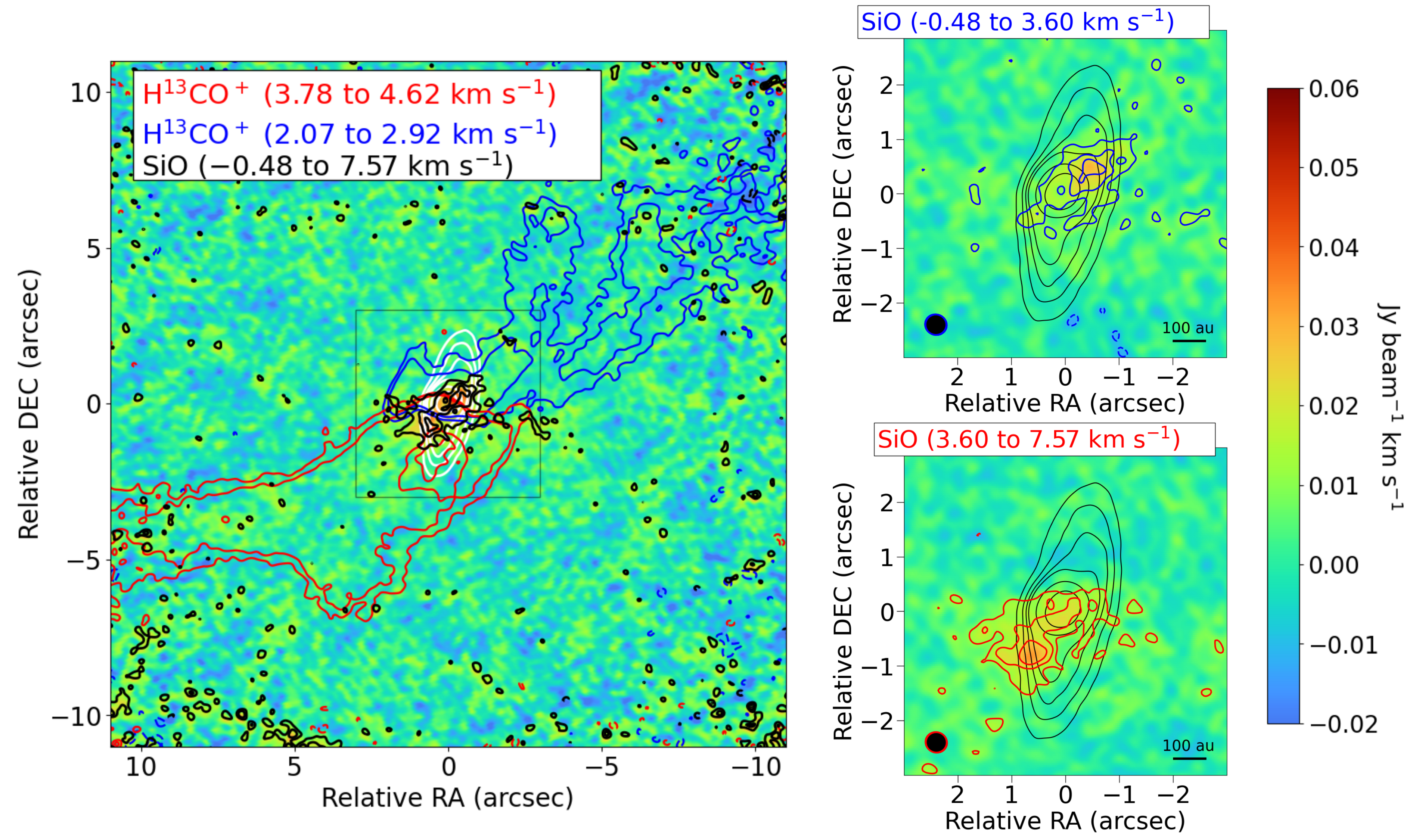}
       \caption{\textit{Left:} SiO moment 0 map (color map and black contours) overlaid with white contours from the continuum and red and blue contours from the moment 0 maps of H$^{13}$CO$^+$. The velocity ranges are indicated in the panels. The SiO contours are [-5,3,5,7] times the rms noise value of 5.0 mJy~beam$^{-1}$~km~s$^{-1}$, with negative contours in dashed lines. The H$^{13}$CO$^+$ contours are [-5,10,15] times the 1$\sigma$ rms noise level of 1.0\,mJy~beam$^{-1}$~km~s$^{-1}$.  Continuum contours are the same as Figure \ref{fig:moment0}. \textit{Right:} Continuum (black contours) and integrated intensity maps of SiO (color map and colored contours) in the area covered by the square in the left panel. Contour levels are the same as left panel, for rms noise values of 3.5\,mJy~beam$^{-1}$~km~s$^{-1}$ for both blueshifted and redshifted integrated SiO emission. The synthesized beam sizes of 0.43$\times$0.39" for SiO and 0.34$\times$0.33" for continuum are represented with the ellipse at the bottom left corners, with the same color as the corresponding contours. The rms noise of the integrated intensity maps has been measured before correcting for the primary beam response. The range of the colorscale has been expanded with respect to Fig. \ref{fig:moment0} for better visualization of the SiO emission.} \label{fig:sio}
    \end{figure*}

    The integrated intensity map of the SiO\,(5-4) transition at 217.105 GHz ($E_u=31.26$\,K) is presented in Figure \ref{fig:sio}. The emission is centered at the peak of the 1.3 mm dust continuum, and has a slightly elongated shape with a size of $\sim3$" and a PA of 114 $\pm$ 6$^{\circ}$ obtained from Gaussian fitting. This orientation deviates $56\pm6^{\circ}$ from the CO outflow axis (PA=$58\pm2^{\circ}$, Sect. \ref{sec:elongations}) and does not either align with the PA of the continuum (167$^{\circ}$, \citealt{alves2017}). The integrated intensity maps show that the SiO emission is aligned with the large-scale structures seen in H$_{2}$CO, H$^{13}$CO$^{+}$, CCH and c-C$_3$H$_2$ in contrast to the CO outflow orientation. Moreover, the redshifted emission extends towards the SE and the blueshifted emission towards the NW, as for the other molecular tracers.

    \subsection{Kinematics}\label{sec:kinematics}

\begin{figure*}[h]
       \centering
       \includegraphics[width=\textwidth]{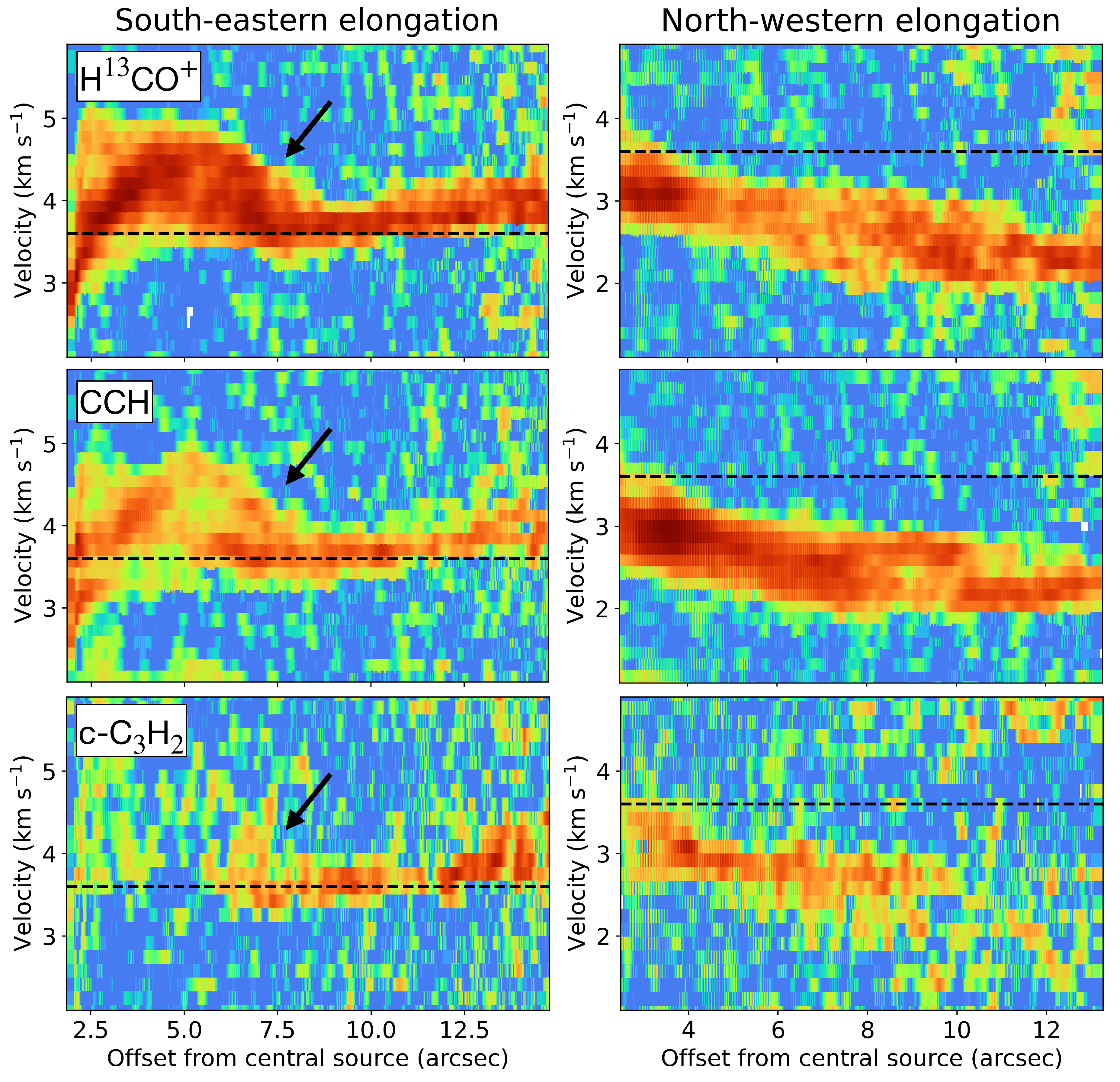}
       \caption{PV diagrams of the H$^{13}$CO$^+$ (top panels), CCH (middle) and c-C$_3$H$_2$ (bottom) emission from the areas highlighted in Figure \ref{fig:moment0}. Left panels show the redshifted emission of the putative second outflow, while right panels show the blueshifted emission counterparts. The dashed lines represent the systemic velocity of the source, $v_{sys}\sim$\mbox{3.6 km~s$^{-1}$}. The arrows in the redshifted panels correspond to the transition between the rotating envelope and the outflowing gas, as also specified in Fig. \ref{fig:moment0}. We have employed a logarithmic colorscale to emphasize the shape of the velocity curve of the outflowing gas.
       }\label{fig:pvsoutflow}
    \end{figure*}

    We investigate the kinematics of the elongated components with the position-velocity (PV) diagrams shown in Figure \ref{fig:pvsoutflow} obtained for the emission enclosed in the regions defined in Figure \ref{fig:moment0}. The PV diagrams of the redshifted emission show two components: from the center up to a distance of $\sim$7.5\,", the velocity profile is consistent with rotation from the flattened envelope \citep{alves2017}. From this radial distance onwards, the velocity curve is either flat (as seen for H$_2$CO, \citealt{evans2023}) or it slightly accelerates when moving away from the systemic velocity. The latter case is a clear sign of outflowing motions from the central source. We have marked the transition region between the rotating envelope and the outflowing gas with arrows, which also corresponds to the change in morphology of the H$^{13}$CO$^+$ and CCH emission, and with the appearance of the outflow cavity tracer c-C$_3$H$_2$ also marked with arrows at the same distance in Fig. \ref{fig:moment0}. The blueshifted velocity curve clearly reveals accelerating outflowing gas without contamination from the rotating envelope, consistent with what was already seen in the PV diagram of the H$_2$CO blueshifted structure \citep{evans2023}.
    
    \subsection{Physical properties}\label{sec:physicalproperties}

    \subsubsection{Mass, momentum and energy} \label{sec:massmomentumenergy}

    From Fig. \ref{fig:pvsoutflow}, we estimate a projected extent of $\sim$13\," or $\sim2000$ au for the elongated structures seen in H$^{13}$CO$^+$, CCH, and c-C$_3$H$_2$, as measured from the central system. This value is a lower limit since the emission is detected up to the borders of the field of view of our observations and more material might be present beyond this point. The radial velocity curves are almost flat or show a very small velocity gradient (Fig. \ref{fig:pvsoutflow}), which makes it difficult to estimate their actual expansion velocity. The largest velocity difference with respect to $v_{sys}$ at which we detect the emission from the elongated H$^{13}$CO$^+$ structure is $1.68\pm0.14$ km~s$^{-1}$ (i.e. 12 pixels in the blueshifted PV diagrams of Fig. \ref{fig:pvsoutflow}, where the velocity resolution is 0.14 km~s$^{-1}$). We assume this to be the expansion velocity $v_{exp}$ of the outflow. However, projection effects must be considered. The inclinations of the circumstellar disks with respect to the line of sight are of $40\pm10^{\circ}$ \citep{alves2019}. Assuming that the outflow is launched from one of the circumstellar disks and that its axis is perpendicular to the circumstellar disk plane, $v_{exp}$ is corrected to $2.2\pm0.4$ km~s$^{-1}$. Correcting the length of the structure by the same inclination, we obtain a physical extent of $3300\pm700$ au for the elongated structures. Hence, the corrected dynamical timescale of the outflow would be \mbox{$t_{dyn}=7000\pm1900$ yr}.

    We estimate the mass of the outflowing gas by computing the column density of the C$^{18}$O\,(2-1) line emission, which is included in the spectral setup of the FAUST observations. To do this, we have obtained the integrated C$^{18}$O line profile across the blue polygon shown in the right panels of Figure \ref{fig:moment0}. For the redshifted emission, we have used the emission within the part of the red polygon east to the arrow and the dashed line shown in the left panels of Figure \ref{fig:moment0} in order to avoid the contribution from the rotating envelope. The Spectral Line Identification and Modelling (SLIM) tool within the MADCUBA package \citep{martin2019} is employed to fit the integrated C$^{18}$O line profile, assuming optically thin emission under LTE conditions and a warm outflow with a range of temperatures of 50 to 100\,K \citep{beuther2002,podio2021}. The derived column density for the blueshifted lobe is $(2.69\pm0.19)\times$10$^{15}$~cm$^{-2} - (4.4\pm0.3)\times$10$^{15}$~cm$^{-2} $ for temperatures of 50\,K and 100\,K, respectively. The redshifted emission yields  $(3.63\pm0.17)\times$10$^{15}$~cm$^{-2} - (5.9\pm0.3)\times$10$^{15}$~cm$^{-2} $. We use these results to estimate the mass of H$_2$ in the outflow as

    \begin{equation}
    \centering
        M_{H_2} = N\textrm{(C}^{18}\textrm{O}\textrm{)} \left[ \frac{\textrm{H}_2}{\textrm{C}^{18}\textrm{O}} \right] \mu_g m_H d^2 A
    \label{eq:mass}
    \end{equation}

    \noindent where $\left[\textrm{H}_2/\textrm{C}^{18}\textrm{O} \right]$ is obtained considering the isotopic ratio $^{16}$O$/^{18}$O$=560\pm25$ \citep{wilson1994} and the relative abundance of CO to H$_2$, $\chi(\textrm{CO})\sim10^{-4}$, $\mu_g=2.3$ is the mean atomic weight of the gas, $m_H$ is the mass of the hydrogen atom, $d=163\pm5$ pc is the distance towards [BHB2007]\,11, and $A$ is the area across which the C$^{18}$O emission has been integrated (21.4 and 24.1\,"$^2$ for the blueshifted and redshifted lobes, respectively). For the blueshifted lobe, we calculate a mass of  $(4.0\pm0.4)\times10^{-3}M_{\odot} - (6.7\pm0.6)\times10^{-3}M_{\odot}$ for temperatures of 50 and 100\,K, respectively; the results for the redshifted lobe are $(5.0\pm0.5)\times10^{-3}M_{\odot} - (8.0\pm0.8)\times10^{-3}M_{\odot}$. Given the small difference between the values at different temperatures, we adopt mean values of $(5.4\pm0.7)\times10^{-3}M_{\odot}$ and $(7.5\pm0.9)\times10^{-3}M_{\odot}$ for the blueshifted and redshifted emission, respectively. This yields a total outflowing mass of \mbox{$M=(1.29\pm0.11)\times10^{-2}M_{\odot}$}. Using the mass of the outflow, the estimated dynamical timescale of \mbox{$t_{dyn}=7000\pm1900$ yr} and the measured expansion velocity of $v_{exp}=2.2\pm0.4$ km~s$^{-1}$, we estimate an outflow mass loss rate of \mbox{$\dot{M}=M/t_{dyn}=(1.8\pm0.5)\times10^{-6}M_{\odot}\textrm{ yr}^{-1}$}, a linear momentum injection rate of \mbox{$\dot{P}=\dot{M}v_{exp}=(4.0\pm1.3)\times10^{-6}M_{\odot}$ km~s$^{-1}$ yr$^{-1}$}, and a kinetic energy of $E=\frac{1}{2}Mv_{exp}^2=(6\pm2)\times10^{41}$ erg. The energy injection rate defined as $\dot{E}=E/t_{dyn}$ is $(9\pm4)\times10^{37}\textrm{erg yr}^{-1}$. These results are summarized in Table \ref{tab:results} and discussed in Sect. \ref{sec:discussion}.

    However, outflow rates derived from the dynamical timescale $t_{dyn}$ may be overestimated since they consider that the whole outflow moves at the maximum velocity, which is, in general, not true. Following the procedure from \cite{hsieh2023}, we have employed their Pixel Flux-tracing Technique (PFT) code to compare the results with the $t_{dyn}$ method. The code first calculates the C$^{18}$O column density for every pixel of the outflow. Then, for every velocity channel of velocity $v$ with respect to the systemic velocity, the crossing time $\Delta R/v$ is computed, where $\Delta R$ is the pixel size. These values are corrected by the inclination of the outflow with respect to the line of sight. This yields maps of the instantaneous outflow rates for every pixel. In Table \ref{tab:tdynvspft} we compare the results between both methods, where the PFT results have also been obtained for temperatures of 50\,K and 100\,K and averaged. The outflow mass is obtained as the sum of the mass of every pixel and is a factor 5 higher than the mass obtained from the column density derived from the fit to the line profile, proving that both methods are comparable. Similarly, the energy of the outflow is comparable between both methods. The mean mass, momentum and energy rates obtained from the PFT method are between one and two orders of magnitude lower than the ones obtained with $t_{dyn}$, which was also found by \cite{hsieh2023}. This illustrates that the outflow rates derived with the $t_{dyn}$ method must be taken with caution.

    \subsubsection{Rotation of the outflow} \label{sec:rotation}

    \begin{figure*}[h]
       \centering
       \includegraphics[width=\textwidth]{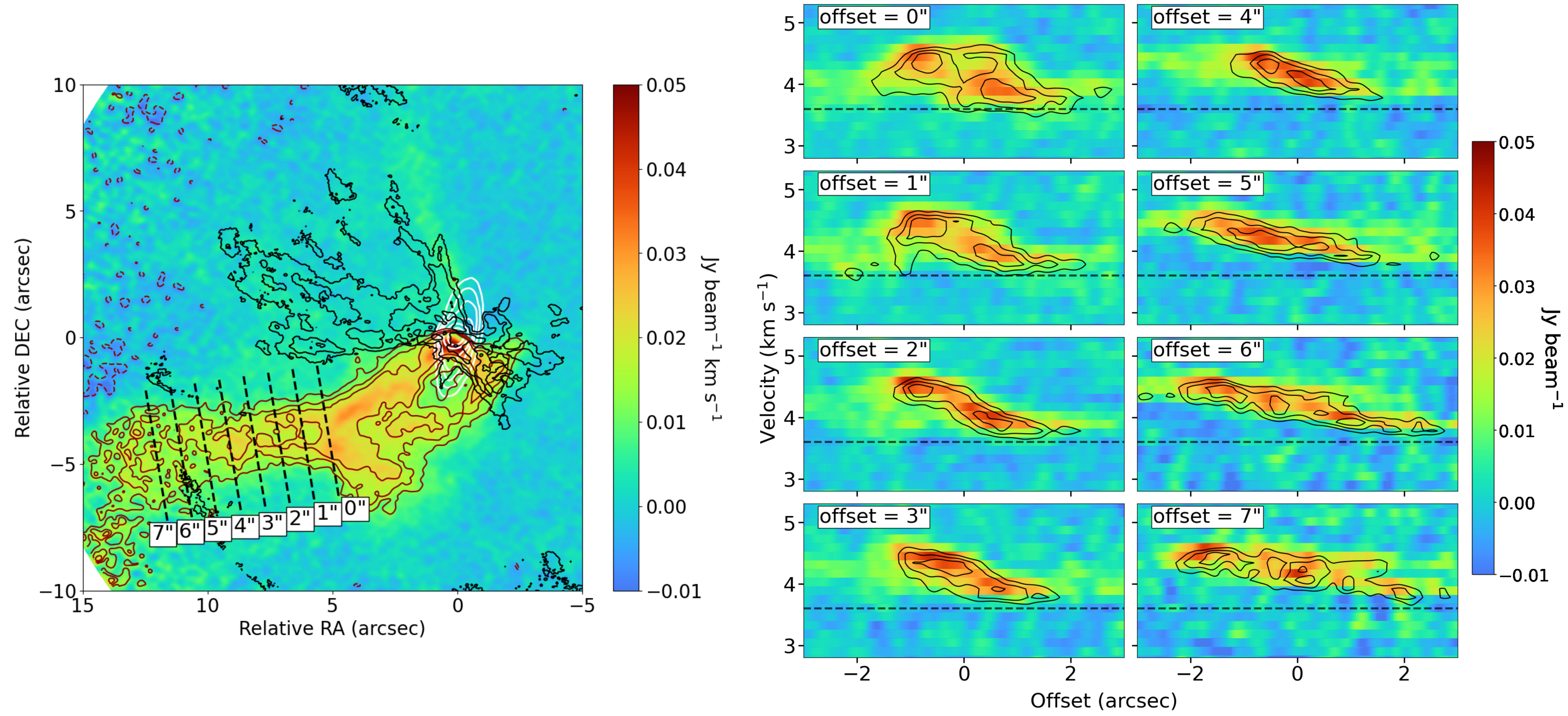}
       \caption{PV diagrams of the H$^{13}$CO$^+$ emission along the transversal direction of the outflow. In the left panel, we show the redshifted H$^{13}$CO$^+$ and CO\,(3-2) integrated intensity maps using the same contours and velocity ranges as in Fig. \ref{fig:moment0}. The dashed lines in the left panel represent the cuts separated by 1\," used to obtain the PV diagrams reported in the right panels. Contours in the PV diagrams are [0.4, 0.6, 0.8] times the peak intensity of the image, which is 0.05 Jy~beam$^{-1}$ for every panel. Negative offsets refer to the northern part of the outflow. Dotted lines in the PV diagrams indicate the systemic velociy of the source, $v_{sys}\sim$\mbox{3.6 km~s$^{-1}$}. Text insets in both panels show the offset of each cut from the rightmost cut, which corresponds to an offset of 0\,".
       }\label{fig:transversalpvs}
    \end{figure*}

    \begin{figure}[h]
       \centering
       \includegraphics[width=\columnwidth]{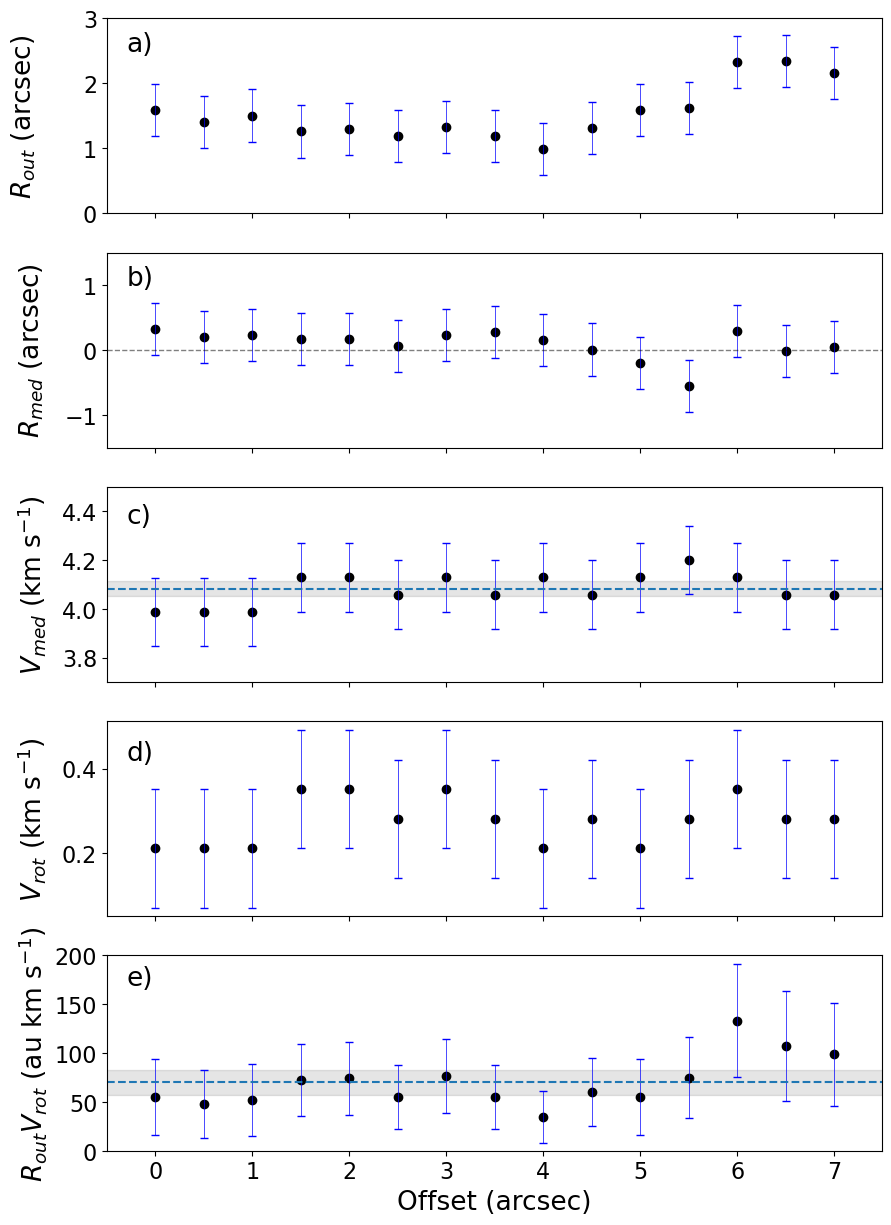}
       \caption{Properties of the rotation of the outflow derived from the transversal PV diagrams of the H$^{13}$CO$^+$ redshifted emission (Fig. \ref{fig:transversalpvs}) at different offsets $z$, with $z$ increasing as one moves away from the central star. $R_{out}$ represents the radius of the outflow, $R_{med}$ is the deviation from the chosen outflow axis, $V_{med}$ is the median line of sight velocity, $V_{rot}$ is the rotation velocity of the outflow and $R_{out}V_{rot}$ is the specific angular momentum of the outflow: all quantities are defined in Sect. \ref{sec:rotation}. The errorbars are computed from the synthesized beam size of $\sim0.4$\," and the spectral resolution of 0.14 km~s$^{-1}$. The dotted line in panel b) corresponds to a value of 0\,". The dotted lines and shadowed areas in panels c) and e) represent the mean value and standard deviation of the points, respectively. }\label{fig:angularmomentum}
    \end{figure}

    \begin{figure}[h]
       \centering
       \includegraphics[width=\columnwidth]{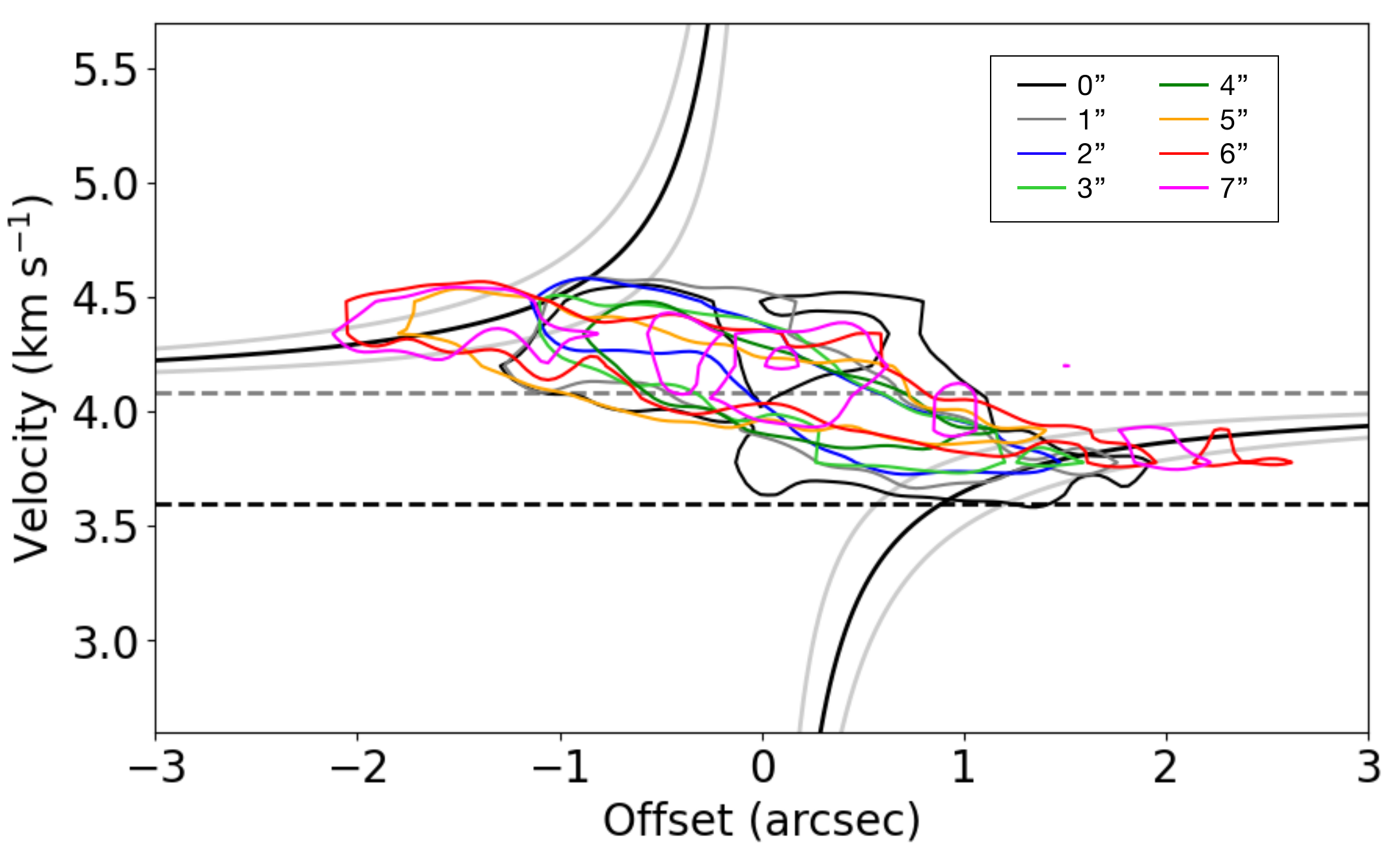}
       \caption{Contours of 0.6 times the maximum value of the PV diagrams from Figure \ref{fig:transversalpvs}. The colors represent different offsets along the outflow, as indicated by the text insets. The black dotted line corresponds to the systemic velocity, $v_{sys}\sim$\mbox{3.6 km~s$^{-1}$}, and the grey dotted line is the mean velocity of the velocity curves, 4.08 km~s$^{-1}$. The black solid curves correspond to a constant specific angular momentum of 70$\pm$25 au~km~s$^{-1}$ for a central velocity of 4.08 km~s$^{-1}$, where the gray curves represent the limits of the error interval. }\label{fig:pvs_angularmomentum}
    \end{figure}

    Following a similar procedure to \cite{zhang2018} and \cite{ohashi2022}, we have explored the transversal motion of the outflow in order to characterize a possible rotation. We focus on the redshifted emission since it has a higher S/N ratio. In Figure \ref{fig:transversalpvs} we show a set of PV diagrams obtained perpendicular to the elongated H$^{13}$CO$^+$ emission with a PA = 8$^{\circ}$. The rightmost cut, closest to the star, corresponds to the PV diagram with an offset of 0\," , which is the point from which the H$^{13}$CO$^+$ emission traces the outflow instead of the rotating envelope (see arrows in Figs. \ref{fig:moment0} and \ref{fig:pvsoutflow}). The PV diagrams reveal a velocity gradient across the axis of the outflow, with the red-shifted gas located toward the north and the blueshifted gas located toward the south, suggesting  either rotation of the gas or shear at the boundary between an outflow and the ambient gas (see Section \ref{sec:alternative}). This gradient has the opposite sense with respect to the NE cone of the CO outflow, which shows blueshifted emission in the north and redshifted gas in the south \citep[][Section \ref{sec:morphology}]{alves2017}. At offsets further away from the central binary system, the velocity curves are more extended and the velocity gradient flattens out, which would be explained by the progressive opening of a conical outflow and slowing down rotation at larger distances from the launching point. This is the same behaviour as those seen in NGC 1333 IRAS 4C \citep{zhang2018} and VLA 1623A \citep{ohashi2022}.

    We then employ the PV diagrams to quantify the rotation of the outflow as done in \cite{zhang2018}. For each PV diagram at an offset $z$, we have obtained the most extreme velocity and distance of the northern and southern parts of the outflow, $v_N(z)$ and $r_N(z)$, and $v_S(z)$ and $r_S(z)$, respectively. These points are obtained at the extremes of the curve for each half of the diagram, i.e. for the northern and southern emission, within the pixels of 0.6 times the value of the peak intensity, which adequately trace the extent of the curve (Fig. \ref{fig:transversalpvs}). With these values, we compute the radius of the outflow, $R_{out}(z)=[r_S(z)-r_N(z)]/2$, the deviation from the chosen outflow axis $R_{med}(z)=[r_S(z)+r_N(z)]/2$, the median line of sight velocity $V_{med}(z)=[v_S(z)+v_N(z)]/2$ and the rotation velocity $V_{rot}(z)=[v_N(z)-v_S(z)]/2$. The results are shown in Figure \ref{fig:angularmomentum}. Panel a) shows that the outflow extension is nearly constant at lower offsets and increases at larger radial distances from the central binary system, i.e. at larger values of $z$ as seen in Figure \ref{fig:transversalpvs}. Panel b) allows us to evaluate how accurate the outflow axis choice for the PV diagrams has been. The points are distributed about 0\,", which indicates a correct estimate. Panel c) shows an almost constant outflow line of sight expansion velocity with a mean value and standard deviation of $4.08\pm0.06$ km~s$^{-1}$. In panel d) we show the rotation velocity of the outflow, i.e. the gradient between the line of sight velocity of the northern and southern parts of the outflow. This panel illustrates one of the limitations of the method with our dataset: the spectral resolution of $\sim0.14$ km~s$^{-1}$ is similar to the velocity gradient of the outflow, resulting in differences of 4 or 5 pixels at most. The error bars, computed from this spectral resolution, are larger than the separation between points, hence these results should be taken with caution. We combine the information from panels a) and d) to obtain the specific angular momentum of the outflow $l(z)=R_{out}(z)V_{rot}(z)$, shown in panel e). The derived mean and standard deviation of the specific angular momentum is $70\pm25$ au~km~s$^{-1}$, which lies within the error bars of the value of $\sim100$ au~km~s$^{-1}$ found for NGC 1333 IRAS 4C \citep{zhang2018} and VLA 1623A \citep{ohashi2022}. The angular momentum is basically constant, and the points follow essentially the shape of the $R_{out}$ distribution in panel a). We note that the values of the specific angular momentum should be taken with caution given the large uncertainty in estimating $V_{rot}$. In Figure \ref{fig:pvs_angularmomentum}, we plot the curve of constant specific angular momentum of 70 au~km~s$^{-1}$ with the 60\% intensity emission contours obtained from the PV diagrams of Fig. \ref{fig:transversalpvs}. The curve encloses the emission from the velocity curves at different offsets, which illustrates that the velocity gradients are consistent with rotation with a constant specific angular momentum.

    Last, we estimate the launching radius $r_0$ of the outflow employing the method from \cite{anderson2003}:

    \begin{equation}
    \centering
        r_0 \approx 0.7 \textrm{au} \left( \frac{l}{100 \textrm{ au~km~s}^{-1}} \right)^{2/3} \left( \frac{v_{exp}}{100 \textrm{ km~s}^{-1}} \right)^{-4/3} \left( \frac{M_*}{1M_{\odot}}\right)^{1/3}
    \label{eq:launchingradius}
    \end{equation}   
    
    \noindent where $l$ is the mean specific momentum and $M_*$ the mass of the central system. We consider the velocity difference of the outflow of $v_{exp}=2.2\pm0.4$ km~s$^{-1}$ (Section \ref{sec:massmomentumenergy}), the combined mass of the system \citep[$2.25\pm0.13$ $M_{\odot}$; see][]{alves2019} and the derived specific angular momentum $l=70\pm25$ au~km~s$^{-1}$. If we assume that the outflow is launched by one of the companions, and considering an equal-mass system as an approximation, the estimated launching radius is $r_0=90\pm30$\,au. This radius is at the scale of the circumbinary disk and is similar to the launching radius of the CO outflow, $\sim90-130$\,au  \citep{alves2017}. However, this estimate is very sensitive to the value of the specific angular momentum and to the expansion velocity of the outflow $v_{exp}$, which is estimated with poor accuracy given the low velocity gradient of the velocity curves (Fig. \ref{fig:pvsoutflow}). A value of $v_{exp}=10$ km~s$^{-1}$, as assumed for NGC 1333 IRAS 4C \citep{zhang2018} and VLA 1623A \citep{ohashi2022}, would result in an estimate of the launching radius of $r_0\approx$10 au, i.e. at the scales of the orbit of the binary system. Note that source B is less massive than its companion and is the main accretor of the system \citep{alves2019}, which is consistent with predictions from simulations \citep[e.g.][]{duffell2020,ceppi2022}. If the outflow is launched by this object, the lower mass would also result in a shorter launching radius than the one derived considering an equal-mass system. 

\begin{table*}[ht!]
        \caption{Properties of the potential new outflow in [BHB2007]\,11. The quantities are corrected from the inclination of the two circumstellar disks, $40\pm10^{\circ}$ \citep{alves2019}.
        } \label{tab:results}
    \begin{center}
    \begin{tabular}{c c c}
         \hline
         Property &  Value & Obtained from \\  
         \hline
         
         Length of a single lobe, $R$  & $3300\pm700$ au & H$^{13}$CO$^+$, CCH and c-C$_3$H$_2$ PV diagrams (Fig. \ref{fig:pvsoutflow}) \\

         Expansion velocity, $v_{exp}$ & $2.2\pm0.4$ km~s$^{-1}$ & H$^{13}$CO$^+$, CCH and c-C$_3$H$_2$ blueshifted PV diagrams (Fig. \ref{fig:pvsoutflow}) \\

         Mass, $M$ & $(1.29\pm0.11)\times10^{-2}M_{\odot}$ & C$^{18}$O column density in the lobes \\

         Dynamical timescale, $t_{dyn}$ & $7000\pm1900$ yr & $t_{dyn}=R/v_{exp}$ \\

        Mass loss rate, $\dot{M}$ & $(1.8\pm0.5)\times10^{-6}M_{\odot}\textrm{yr}^{-1}$ & $\dot{M}=M/t_{dyn}$ \\

         Momentum injection rate, $\dot{P}$ & $(4.0\pm1.3)\times10^{-6}M_{\odot}$ km~s$^{-1}$ yr$^{-1}$ & $\dot{P}=\dot{M}v_{exp}$ \\

         Kinetic energy, $E$ & $(6\pm2)\times10^{41}$ erg & $E=\frac{1}{2}Mv_{exp}^2$ \\

         Energy injection rate, $\dot{E}$ & $(9\pm4)\times10^{37}\textrm{erg yr}^{-1}$ & $\dot{E}=E/t_{dyn}$ \\

         Specific angular momentum, $l$  & $70\pm25\textrm{ au km s}^{-1}$ & H$^{13}$CO$^+$ transversal PV diagrams \\

            &   & in the redshifted lobe (Figs. \ref{fig:transversalpvs}, \ref{fig:angularmomentum} and \ref{fig:pvs_angularmomentum}) \\

        Launching radius, $r_0$  & $90\pm30$ au & H$^{13}$CO$^+$ transversal PV diagrams \\

            &   & in the redshifted lobe (Figs. \ref{fig:transversalpvs}, \ref{fig:angularmomentum} and \ref{fig:pvs_angularmomentum}) \\
        
        SiO abundance, $\chi(\textrm{SiO})$ & $\geq(0.11-2.0)\times10^{-9}$ & SiO, C$^{18}$O and H$^{13}$CO$^+$ column density  \\

            &   & in the SiO emitting region (Fig. \ref{fig:sio}) \\ \hline
                
\end{tabular}
\end{center}
\end{table*}

 \begin{table*}[h]
        \caption{Comparison between the outflow properties obtained from the $t_{dyn}$ method (Table \ref{tab:results}) and from the PFT method \citep{hsieh2023}. Note that mass and energy are not actually computed with $t_{dyn}$ or the PFT crossing time, but we compare them to illustrate that the PFT code yields similar results.} \label{tab:tdynvspft}
    \begin{center}
    \begin{tabular}{c c c c}
         \hline
         Property &  Value from $t_{dyn}$ method & Value from PFT method & Ratio between $t_{dyn}$ and PFT methods \\  
         \hline
         
         Mass, $M$ & $1.29\times10^{-2}M_{\odot}$ & $7.0\times10^{-2}M_{\odot}$ & 0.2 \\

         Mass loss rate, $\dot{M}$ & $1.8\times10^{-6}M_{\odot}\textrm{yr}^{-1}$ & $4.2\times10^{-8}M_{\odot}\textrm{yr}^{-1}$ & 40 \\

         Momentum injection rate, $\dot{P}$ & $4.0\times10^{-6}M_{\odot}$ km~s$^{-1}$ yr$^{-1}$ & $4.4\times10^{-8}M_{\odot}$ km~s$^{-1}$ yr$^{-1}$ & 90 \\

         Kinetic energy, $E$ & $6\times10^{41}$ erg & $1.2\times10^{41}$ erg & 5 \\

         Energy injection rate, $\dot{E}$ & $9\times10^{37}\textrm{erg yr}^{-1}$ & $5.1\times10^{35}\textrm{erg yr}^{-1}$ & 180 \\

         \hline
                
\end{tabular}
\end{center}
\end{table*}

    \subsection{Shocked material at smaller scales}\label{sec:sio} 
    
    To estimate the abundance of the SiO emission, we first obtain the SiO and C$^{18}$O column densities in the region defined by the 3$\sigma$ contour from Figure \ref{fig:sio}, and then use the following expression:

    \begin{equation}
    \centering
        \chi(\textrm{SiO})=\frac{N\textrm{(SiO)}}{N\textrm{(C}^{18}\textrm{O}\textrm{)}}\frac{^{18}\textrm{O }}{^{16}\textrm{O }}\chi(\textrm{CO})
    \label{eq:sio_abundance}
    \end{equation}

    \noindent with the same values of $^{16}$O$/^{18}$O and $\chi(\textrm{CO})$ used in Sect. \ref{sec:physicalproperties}. To calculate the column density of C$^{18}$O and SiO, we have considered LTE and optically thin emission:

    \begin{equation}
    \centering
        N_T = \frac{N_u}{g_u}Q(T_{ex})e^{E_u/kT_{ex}}
    \label{eq:columndensity}
    \end{equation}

    \noindent where $N_T$ is the total column density of the molecule, $N_u$ is the column density of molecules in the upper energy level, $g_u$ is the statistical weight of the upper level, $Q(T_{ex})$ is the partition function at the excitation temperature $T_{ex}$, and $E_u$ is the upper level energy. To obtain $N_u/g_u$, we use equation (12) from \cite{dierickx2015}:

    \begin{equation}
    \centering
        \frac{N_u}{g_u}=1.669\times10^{17}\frac{\int T_B \,dv\, \textrm{(K km s$^{-1}$)}}{S\mu\textrm{(D)}^2\nu\textrm{(MHz)}}\,\textrm{cm}^{-2}
    \label{eq:nugu}
    \end{equation}

    \noindent where $\int T_B \,dv$ is the integrated line profile, $S$ is the line strength and $\mu$ is the dipole moment. The line profile of both molecules is obtained from their respective data cube with a spatial aperture defined by the 3$\sigma$ contour from Figure \ref{fig:sio}. We integrate the line profiles avoiding the central $\pm$3 km~s$^{-1}$ velocities from the systemic velocity of 3.6 km~s$^{-1}$ to minimize the effect of C$^{18}$O tracing the rotating envelope. 
    
    Since only one SiO transition is covered by our setup, we assume a range of temperatures between 50 and 100~K in our calculations, given the temperature gradients observed in protostellar outflows \citep{beuther2002,podio2021}. Equations (\ref{eq:columndensity}) and (\ref{eq:nugu}) in the selected velocity ranges yield column densities of $N(\textrm{SiO})=(3.4-5.0)\times10^{13}$cm$^{-2}$ and $N\textrm{(C}^{18}\textrm{O}\textrm{)}=(4.7-7.9)\times10^{16}$cm$^{-2}$ for a temperature range of 50$-$100~K. From equation (\ref{eq:sio_abundance}), this results in abundances of SiO of $\chi(\textrm{SiO})\geq (1.1-1.3)\times10^{-10}$ with respect to H$_2$.

    Another possibility is to estimate the SiO abundance with the SiO and H$^{13}$CO$^+$ column densities:

    \begin{equation}
    \centering
        \chi(\textrm{SiO})=\frac{N\textrm{(SiO)}}{N\textrm{(H}^{13}\textrm{CO}^{+}\textrm{)}}\frac{^{13}\textrm{C }}{^{12}\textrm{C }}\chi(\textrm{HCO}^+)
    \label{eq:sio_abundance_h13cop}
    \end{equation}

    \noindent where $^{12}$C$/^{13}$C$\sim69$ \citep{milam2005} and $\chi(\textrm{HCO}^+)\sim10^{-8}$ \citep{irvine1987}. The H$^{13}$CO$^+$ column density of $(2.5-3.8)\times10^{12}$cm$^{-2}$ yields SiO abundances of $\chi(\textrm{SiO})\geq (1.9-2.0)\times10^{-9}$.
    
    The derived abundances are two to four orders of magnitude lower than the typical values of $10^{-7}-10^{-6}$ found for Class 0 sources that drive collimated jets \citep{podio2021}. Although we have avoided on purpose the central radial velocities, a fraction of the measured C$^{18}$O and H$^{13}$CO$^+$ emission may still arise from the envelope. These central velocities are where most of the SiO line emission is concentrated, while the line wings are much fainter. This inevitably results in an underestimate of the SiO abundance. Moreover, if the SiO emission arises from compact shocks, it is likely that this emission gets diluted within the synthesized beam of our observations and therefore, the SiO column densities that we obtain are probably lower than the actual values. Hence, the derived SiO abundances should be considered as lower limits, which may explain the low values in comparison to other sources. 

%-----------------------------------------------------------------

\section{Discussion}\label{sec:discussion}

    The analysis of multiple molecular tracers has revealed a very detailed picture of the different kinematic components at intermediate scales around [BHB2007]\,11. In the following sections, we discuss the new outflow hypothesis based on the evidence gathered from the different molecular tracers, as well as alternative explanations such as the possibilities of a wide-angle outflow cavity, a relic outflow or accretion from infalling streamers.

    \subsection{A new outflow in [BHB2007]\,11} \label{sec:newoutflow}

    \begin{figure*}[h]
    \centering
    \includegraphics[width=\textwidth]{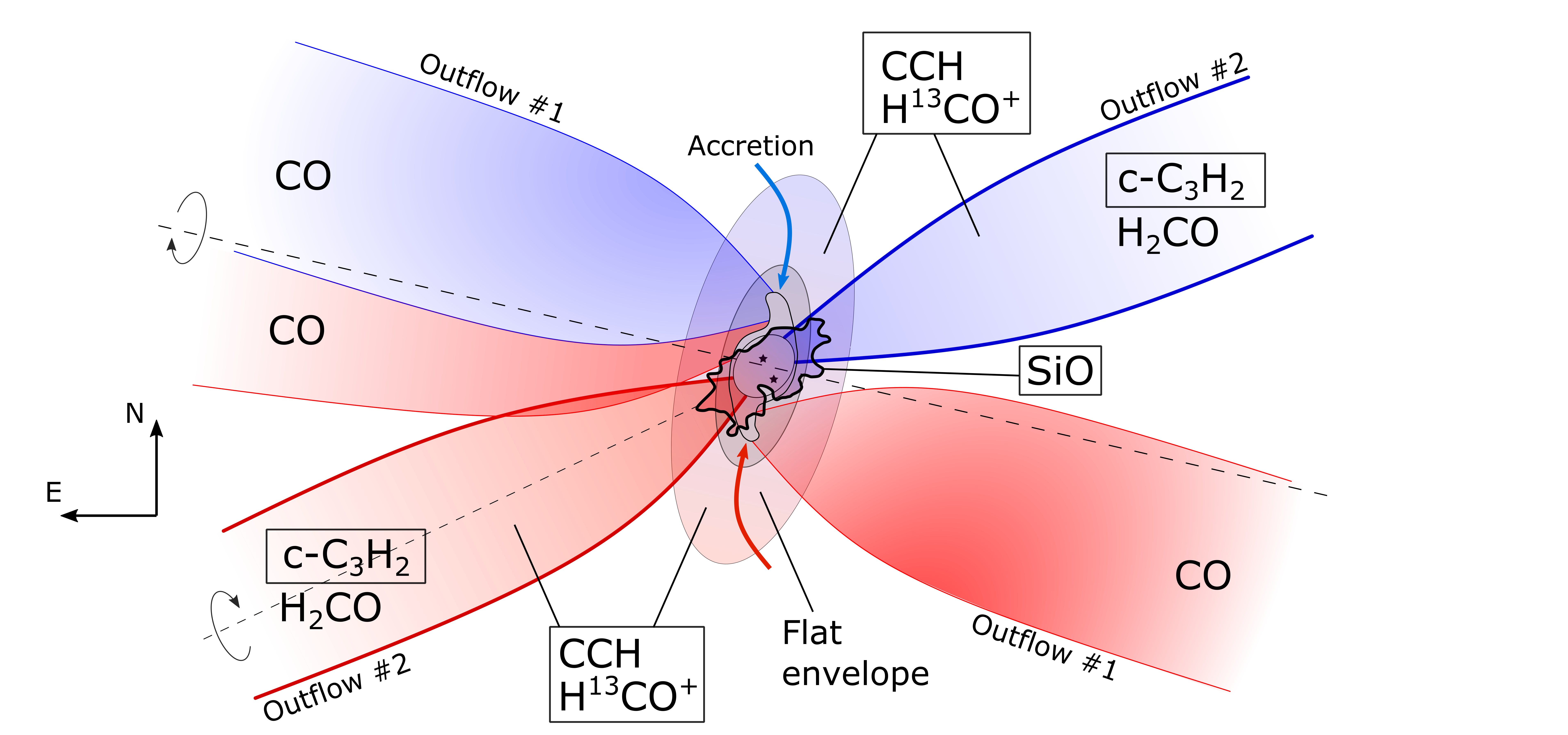}
    \caption{Sketch - not to scale - of the proposed interpretation of the observations towards [BHB2007]\,11 where a second outflow is found in addition to the CO outflow reported by \citet{alves2017}. The ellipses at the center represent the flattened envelope. The innermost contour represents the circumbinary disk, where the two stars of the system are represented by the star symbols. The inner ellipse represents the continuum contours from Figures \ref{fig:moment0} and \ref{fig:sio}, where the spiral structures seen in \cite{alves2017} are emphasized. The tips of these spiral structures are the landing sites of the accretion from the envelope (blue and red arrows) and the launching points of the CO outflow \citep{alves2017}. We have marked the CO outflow from \cite{alves2017} as 'Outflow \#1', while the new proposed outflow inferred from the observations in this work is named 'Outflow \#2'. The launching point of Outflow \#2 is unknown and we draw it from the edge of the circumbinary disk. The SiO emission is represented with the thick contour at the center, corresponding to the 3$\sigma$ contour from the left panel of Figure \ref{fig:sio}. The different molecular tracers used to identify each component of the system are overplotted to the sketch in the corresponding part. The rotation axes are indicated with the dashed lines. Blue and red colors represent blueshifted and redshifted velocities, respectively.}\label{fig:sketch}
    \end{figure*}
    
    Our main hypothesis derived from the emission of the multiple molecular lines analyzed in this work consists of a second outflow in addition to the main CO outflow reported by \cite{alves2017}. In Figure \ref{fig:sketch} we present a summary of the proposed picture.
    
    The structures discovered in H$_2$CO by \cite{evans2023} have been identified in H$^{13}$CO$^+$, CCH and c-C$_3$H$_2$ (Figure \ref{fig:moment0}). The PV diagrams obtained in these structures (Figure \ref{fig:pvsoutflow}) are consistent with outflowing motion in the blueshifted part, while the redshifted lobe shows tentative acceleration although not as steep as for the blueshifted gas. The velocity curves are similar to those obtained from H$_2$CO by \cite{evans2023}, which could not be fitted by infalling streamer modeling \citep{pineda2020}. These results suggest that the two elongated structures are lobes of a second outflow powered by [BHB2007]\,11, which remained unidentified in previous works. This was due to the fact that this second outflow may lie close to the plane of the sky and therefore, its velocities lie in a velocity range that is largely affected by self-absorption in the observed CO line emission, since CO is also present in the extended envelope \citep{alves2017}. The fact that c-C$_3$H$_2$ is only seen in the elongated structures and at the base of the CO outflow reinforces the second outflow hypothesis, as this molecule has been observed in UV-illuminated environments such as outflow cavities \citep[Sect. \ref{sec:elongations}]{murillo2018,tychoniec2021}. However, while the blueshifted lobe clearly shows acceleration motions, the redshifted lobe might be perturbed by a wider cavity caused by a slower component of the CO outflow (Sect. \ref{sec:cavity}). It is possible that both structures coexist and emit in the same region of the sky, complicating the interpretation of the redshifted emission with the current dataset. Additionally, the velocity gradient seen in the blueshifted emission may be caused by a bending of the outflowing gas along the direction of the line of sight. To test this hypothesis, we have estimated the velocity difference that would result from a deflection of the outflow. For an initial angle of the outflow with respect to the line of sight of $\theta=40^{\circ}$, i.e. the same orientation as the circumstellar disks \citep{alves2019}, a minimum outflow expansion velocity $v_{exp}$ of 7.3 km~s$^{-1}$ is needed to cause a gradient of 1.7 km~s$^{-1}$ (Fig. \ref{fig:pvsoutflow}) purely by bending, and it would require a bending of $40^{\circ}$ towards the line of sight. However, in this case, the minimum observed velocity difference with respect to the systemic velocity would be of 5.6 km~s$^{-1}$, which is much faster than the velocities that we detect. If the initial orientation of the outflow were closer to the plane of the sky, lower outflow velocities would be needed to reproduce the velocity gradient, but conversely the deflection angles would be much higher. This suggests that the gradient cannot be explained by a bending of the outflow.
    
    Table \ref{tab:results} collects the properties of the second outflow derived in Section \ref{sec:results}. The derived mass loss rate of $(1.8\pm0.5)\times10^{-6} M_{\odot}\textrm{ yr}^{-1}$ is among the typical values of jets from Class 0 and Class I sources \citep[e.g.][]{bally2016,lee2020,podio2021}, and this rate decreases as the protostar enters the Class I phase. The recent survey study from \cite{hsieh2023}, who obtain outflow properties with CO isotopologues observed with ALMA, yield higher values for the mass and mass loss rate of the outflows, which are within a factor of two times higher than our derived values. As we have noted in Section \ref{sec:physicalproperties}, one of the main limitations of our data is that the outflow is potentially not fully covered by the field of view of the observations. This implies that properties such as the mass and the extent of the outflow are underestimated. In addition, the expansion velocity of the outflow $v_{exp}$ is potentially neither accurately constrained since it may expand close to the plane of the sky, given the small gradients in the velocity curves from Fig. \ref{fig:pvsoutflow}. Hence, quantities involving the expansion velocity, such as the mass loss and momentum rates and the energy of the outflow are also potentially underestimated, yielding lower values than those found in the literature. We note that our calculations consider a single expansion velocity for the whole outflow, which is an approximation that may not account for the true velocity of the entrained material. 

    The PV diagrams obtained along the transversal direction of the redshifted lobe (Fig. \ref{fig:transversalpvs}, Sect. \ref{sec:rotation}) have revealed transversal motions in the opposite sense to the rotation of the CO outflow \citep{alves2017}. Assuming that the transversal motion corresponds to rotation of the second outflow, an analysis following the methodology of \cite{zhang2018} yields a specific angular momentum of $70\pm25$ au km s$^{-1}$. However, this method is very limited in the case of [BHB2007]\,11 due to the low velocity gradients that are comparable to the velocity resolution of the ALMA observations (see panel d) in Figure \ref{fig:angularmomentum}). This, in addition to the poorly constrained expansion velocity of the outflow, prevents us from strongly constraining the outflow launching radius with the current dataset. The derived value of the launching radius is $90\pm30$\,au, which corresponds to the scales of the circumbinary disk, although slightly higher expansion velocities would result in much shorter launching radii (Sect. \ref{sec:rotation}). A value of $v_{exp}=10$ km~s$^{-1}$, as assumed for NGC 1333 IRAS 4C \citep{zhang2018} and VLA 1623A \citep{ohashi2022}, would result in a launching radius of $r_0\approx$10 au. 
    
    The SiO emission detected at the inner $\sim300$ au extends along the direction of the large-scale structures seen in H$_{2}$CO, CCH, H$^{13}$CO$^{+}$ and c-C$_3$H$_2$, with the redshifted emission seen towards the SE and the blueshifted emission towards the NW (Figure \ref{fig:sio}). SiO is produced as a consequence of the sputtering of dust grains in shocks \citep{martinpintado1992,caselli1997,schilke1997,jimenezserra2004}, which makes of it an excellent outflow shock tracer. The orientation of the CO outflow is almost perpendicular to the detected SiO structure, which leads us to rule out the possibility of SiO tracing the base of the CO outflow. In addition to this, the observed SiO emission peaks toward the dust emission peak (and hence its origin is located within the central $\sim0.4$\,"), which contrasts with the large launching radius of the CO outflow of $\sim$90-130 au (i.e. at more than 0.55\,") from the center. The abundance of SiO within the structure is $\geq(0.11-2.0)\times10^{-9}$. Grain sputtering models show that shock velocities between 10 and 20 km~s$^{-1}$ are needed to sputter sufficient silicon from the icy mantles of dust grains to yield SiO gas-phase abundances $\sim$10$^{-9}$ \citep[see e.g.][]{jimenezserra2008,gusdorf2008b,nguyenluong2013}. The most extreme velocity values of the SiO\,(5-4) line emission in our dataset are of 10 km~s$^{-1}$, or $13\pm3$ km~s$^{-1}$ when we correct by the disk inclination of $40^{\circ}\pm10^{\circ}$, which is consistent with the grain sputtering model predictions.
    
    Ejection of matter is tightly tied to accretion processes in young protostars \citep[e.g.][]{bally2016}. There are examples of binary protostars hosting multiple outflows, for instance L1551 IRS5 \citep[e.g.][]{rodriguez2003}, IRAS 16293A \citep[e.g.][]{vanderwiel2019}, \mbox{or the under-debate case of VLA 1632A} \citep{hsieh2020,hara2021,ohashi2022}. Previous works on [BHB2007]\,11 have revealed that accretion is taking place at the flattened envelope scale \citep{alves2017} and in the inner parts of the circumbinary disk \citep{alves2019}. Specifically, the less massive member B is thought to be the main accretor of the binary given the evidence of gas streamers in its surroundings, also in agreement with predictions from simulations \citep[e.g.][]{duffell2020,ceppi2022}. The hypothesis of an additional outflow in addition to the CO outflow launched from the circumbinary disk is consistent with the finding of accretion at the circumstellar disk scale \citep{alves2019,vastel2022}, which must entail angular momentum removal. However, the angular resolution of the current ALMA observations is not high enough to resolve the launching region of the second outflow detected in [BHB2007]\,11, and estimates from the current dataset yield high uncertainties (Sect. \ref{sec:rotation}). Future observations at higher angular resolution may be able to resolve the inner parts of the outflow and provide a reliable estimate of the new outflow launching radius.

    Finally, we note that if the SiO emission is indeed tracing an outflow from one of the two sources from the binary system, and assuming that the outflow is perpendicular to the disk plane, it would imply that the circumstellar disk of the driving source would be misaligned with respect to the circumbinary disk. Misaligned outflows in multiple systems are a frequent outcome of the turbulent fragmentation formation scenario of the binary \citep[e.g.][]{offner2010,offner2023}, in contrast to the disk fragmentation theory in which we expect the circumstellar disk to be aligned with the circumbinary material \citep{bonnellbate1994}. The counter-rotation of the putative new outflow with respect to the CO outflow is also consistent with the turbulent formation scenario. Accretion of material with different net angular momentum causes the misalignment of the circumstellar disks with respect to the axis of the circumbinary disk \citep{bate2018}, which is possible in the turbulent accretion scenario, as well as in asymmetric accretion coming from infalling streamers (Sect. \ref{sec:streamers}).

    \subsection{Alternative interpretations}\label{sec:alternative}
    
        \subsubsection{Cavity walls of the CO outflow} \label{sec:cavity}
        
        Alternatively, the elongated structures might be tracing the walls of the cavity carved by a slower, wide-angle component of the CO outflow \citep[e.g.][]{devalon2022}. The location of the elongations, with the redshifted emission located toward the southeast and the blueshifted emission toward the northwest, is consistent with the rotation of the higher velocity CO outflow. However, under this scenario, one would expect to see both sides of the cavity wall for each lobe, delineating a biconical shape in each side of the outflow. This is the case of the outflow cavity in VLA 1632A traced by CCH \citep{ohashi2022}. The data presented here do not show the expected biconical counterpart of the cavity wall on the other side of the rotation axis of each lobe, although this could be explained by different physical conditions of the ambient cloud.

        In addition, in Section \ref{sec:rotation} we have reported a transversal velocity gradient for the redshifted part of the elongated SE structure. This gradient is interpreted as rotation is the opposite sense to the one observed for the main CO outflow and for the envelope. However, we note that this interpretation needs to be taken with caution: there are signs of transversal motions in molecular outflows that could be mistaken for rotation \citep{frank2014}. Instead of rotation from a separate outflow, this gradient may be explained by shear of the wide-angle wind at the edges of the cavity wall. If this is the correct interpretation, our estimate of the launching radius of the emission should not be considered valid as it relies on the assumption of a rotating outflow, although the derived value is consistent with the one found for the main CO outflow \citep{alves2017}. Another possible interpretation is that the main CO outflow is in Keplerian rotation and the emission traces the outer part of the outflow's rotation curve, where slower velocities are expected for wider distances from the rotation axis.

        To further analyze the possibility of a cavity from a wide-angle wind, Figure \ref{fig:cschannelmaps} in the Appendix presents the channel maps of the CS\,(5-4) transition at 244.935 GHz, also covered by the FAUST spectral setup. CS is a dense gas tracer, and the (5-4) transition emits in the outflow cavity walls \citep[e.g.][]{ohashi2022} and does not show absorption close to the systemic velocity. The CS\,(5-4) emission shows simultaneously the two large scale components that have been discussed: it clearly delineates the cavities of the main CO outflow and it also shows emission over the two elongated structures detected toward the SE and NW. As shown in Figure \ref{fig:cschannelmaps}, the panels at 4.53 and 4.08 km~s$^{-1}$ show the redshifted elongation separated from the wall of the northern lobe of the CO outflow, as well as strong emission near the base of the proposed new outflow. Similarly, the blueshifted emission at 2.74 and 2.44 km~s$^{-1}$ trace the elongated emission toward the NW, and strong emission in the SE-NW direction at the base. While the walls of the CO outflow and the elongated structures appear separated in the plane of the sky, these images alone do not allow us to discern between the new outflow scenario or the case of a wide-angle, slow molecular wind enclosing the main CO outflow seen at higher velocities.
        
       The SiO emission seen along the main axis of the new outflow, however, remains difficult to explain under this scenario. If the features in H$^{13}$CO$^+$, CCH and c-C$_3$H$_2$ are part of a wide-angle, slow molecular wind, the SiO emission should appear aligned in the direction of the main CO outflow (i.e. in the NE-SW direction), which is not observed. As mentioned in Sect. \ref{sec:newoutflow}, a possible explanation is that both structures (the new second outflow and a wide-angle, slow wind) coexist and their emission overlaps toward the south-eastern part of the outflow.

        \subsubsection{Relic outflow} \label{sec:relic}

        The observed elongations may be a relic outflow emission from a past episode of ejection of material. This scenario was proposed by \cite{okoda2021} as an explanation to the elongated emission found in IRAS\,15398-3359 that was perpendicular to the main outflow. The observed low expansion velocity of the elongated structures in [BHB2007]\,11 of $2.2\pm0.4$\,km~s$^{-1}$ would correspond to a velocity much lower than the one from the original ejection episode. However, we note that the CO outflow is launched from the circumbinary disk, not by one of the members of the close binary. This means that if the elongations are part of a past ejection of material from this main outflow, the circumbinary disk should have reoriented  $56\pm6^{\circ}$ since then (Sect. \ref{sec:siomorphology}). SPIRE data at 250 $\mu$m show the dust cavity carved by the main CO outflow extending up to $\sim$30000\,au \citep{sandell2021}, suggesting that the circumbinary disk from which the main outflow is launched has not suffered a recent reorientation. Moreover, \cite{okoda2021} found an arc-like structure of shocked material a few thousand au from IRAS\,15398-3359 traced by SO, SiO and CH$_3$OH following the orientation of the additional elongated emission. The observations explored in this work, however, only show shocked material from the SiO emission at the base of the putative new outflow, with a similar orientation to its large-scale structure seen in the other molecular tracers. Consequently, we consider the possibility of a relic emission from the main CO outflow as unlikely.
        
        \subsubsection{Accretion streamers} \label{sec:streamers}

        One last possibility is that the large-scale structures seen in H$^{13}$CO$^+$, CCH and c-C$_3$H$_2$, and the compact SiO emission, are produced by a gas streamer infalling almost parallel to the plane of the sky, releasing SiO at the interaction region with the circumbinary disk forming an accretion shock. As already mentioned, the free-falling streamer models could not reproduce either the velocity curves of H$_2$CO \citep{evans2023} or the large-scale kinematics of the H$^{13}$CO$^+$, CCH and c-C$_3$H$_2$ gas. We acknowledge the fact that the redshifted part of the new elongated structures seen in these molecular tracers presents an almost flat velocity structure, which could contradict the hypothesis of an expanding outflow. However, the presence of SiO, a molecule that is sputtered from dust grains in shocks \citep[see e.g.][]{caselli1997,jimenezserra2008} which is aligned with the large-scale H$^{13}$CO$^+$, CCH and c-C$_3$H$_2$ structures supports the idea of a new outflow. 
        
        We can estimate the impact velocity of a free-falling streamer of material considering the free-fall velocity (\ref{eq:vff}) and the velocity gradient (\ref{eq:gradvff}) \citep{stahlerpalla2004}:

        \begin{equation}
        \centering
            v_{ff}=-\sqrt{\frac{2GM_*}{R}}
        \label{eq:vff}
        \end{equation}

        \begin{equation}
        \centering
            \nabla v_{ff}=\sqrt{\frac{GM_*}{2R^3}}
        \label{eq:gradvff}
        \end{equation}

        \noindent where $M_*$ is the central mass of the system, $2.25\pm0.13$ $M_{\odot}$ \citep{alves2019}, $R$ is the distance to the tip of the elongated structure, $3300\pm700$ au (Sect. \ref{sec:kinematics}, Table \ref{tab:results}) and $G$ is the gravitational constant. The free-fall velocity at the tip of the streamer is $v_{ff}=1.11\pm0.12$ km~s$^{-1}$, and the velocity gradient is $\nabla v_{ff}=34\pm11$ km~s$^{-1}$~pc$^{-1}$, which yields an impact velocity of the material of $1.6\pm0.2$ km~s$^{-1}$. Moreover, magneto-hydrodynamical simulations predict that infall velocities at scales of 100-1000\,au could be slower than the free-fall velocities \citep{hirano2025}. Shock velocities of less than 2 km~s$^{-1}$ are unlikely to release SiO abundances of 10$^{-9}$ \citep{jimenezserra2008,nguyenluong2013}. \cite{kido2023} interpreted the detection of SiO emission spatially coincident with the landing point of a streamer toward IRAS 16544–1604, with an accretion shock. However, they do not report a value for the measured SiO abundance. In addition, SiO is not predicted to be released in such environments by current models \citep[e.g.][]{vangelder2021}, which seems to rule out the accretion streamer scenario for the SiO emission in [BHB2007]\,11.

%--------------------------------------------------------------------

\section{Conclusions}

In this work, we have analyzed observations of H$^{13}$CO$^+$, CCH and c-C$_3$H$_2$ and SiO obtained towards the young binary [BHB2007]\,11 observed within the FAUST ALMA large program, which have revealed crucial to disentangle the different components and the kinematics of the system at intermediate spatial scales. The molecular emission of H$^{13}$CO$^+$, CCH and c-C$_3$H$_2$ reveals two elongated structures at scales of $\sim$1000 au previously discovered in H$_2$CO \citep{evans2023}. The axis of these structures is almost perpendicular to the previously identified CO outflow launched from the circumbinary disk at $\sim$90-130 au from the center of the binary system \citep{alves2017}. The kinematic analysis of the large-scale gas shows hints of outflowing motions and outflow rotation in the opposite sense to the main CO outflow. The emission from H$^{13}$CO$^+$, CCH and c-C$_3$H$_2$ probes a velocity range of $\pm2$\,km~s$^{-1}$ with respect to the systemic velocity of [BHB2077]11, which could not be probed in previous CO observations due to severe self-absorption of the line \citep{alves2017}. At the central $\sim$300 au, the FAUST images detect compact SiO emission elongated in the same direction and expanding in the same sense as the large-scale structures detected in H$^{13}$CO$^+$, CCH and c-C$_3$H$_2$, suggesting that SiO is tracing the shocked material at the base of the new outflow. The center of the SiO emission coincides with the peak of the dust emission, which indicates that the new outflow is likely powered by one of the two protostars orbiting inside the circumbinary disk. The binary is known to be strongly accreting through gas streamers connected to the circumbinary disk \citep{alves2019}. Hence the ejection of a jet that removes the excess of angular momentum is expected. 

All the evidence from the observations analyzed in this work suggests the presence of a second outflow in [BHB2007]\,11. However, we have explored alternative explanations, namely a wide-angle cavity carved by a slower component of the main outflow, a relic outflow, and the presence of accretion streamers. From the current data, we cannot completely discard the possibility of the coexistence of these structures with an additional outflow. Further high angular resolution observations will cast light onto this complex system. This work illustrates the importance of observing different molecular tracers to explore the morphology and kinematics of jets powered by young protostars, which still remain poorly known.

\begin{acknowledgements}
A.M.-H. and I.J.-.S acknowledge funding from grant PID2022-136814NB-I00 funded by the Spanish Ministry of Science, Innovation and Universities/State Agency of Research MICIU/AEI/ 10.13039/501100011033 and by “ERDF/EU”. A.M.-H. has received support from grant MDM-2017-0737 Unidad de Excelencia "María de Maeztu" Centro de Astrobiología (CAB, CSIC-INTA) funded by MCIN/AEI/10.13039/501100011033. LP, ClCo, EB, and GS acknowledge the PRIN-MUR 2020 BEYOND-2p (Astrochemistry beyond the second period elements, Prot. 2020AFB3FX), the project ASI-Astrobiologia 2023 MIGLIORA (Modeling Chemical Complexity, F83C23000800005), the INAFGO 2023 fundings PROTO-SKA (Exploiting ALMA data to study planet forming disks: preparing the advent of SKA, C13C23000770005),  the INAFGO 2024 fundings ICES (tracking the history of ices from the cradles of planets to comets), the INAF MiniGrant 2022 “Chemical Origins” (PI: L. Podio), and the INAF-Minigrant 2023 TRIESTE (“TRacing the chemIcal hEritage of our originS: from proTostars to plan- Ets”; PI: G. Sabatini). LP, ClCo, EB, and GS also acknowledge financial support under the National Recovery and Resilience Plan (NRRP), Mission 4, Component 2, Investment 1.1, Call for tender No. 104 published on 2.2.2022 by the Italian Ministry of University and Research (MUR), funded by the European Union – NextGenerationEU-Project Title 2022JC2Y93 Chemical Origins: linking the fossil composition of the Solar System with the chemistry of protoplanetary disks – CUP J53D23001600006 – Grant Assignment Decree No. 962 adopted on 30.06.2023 by the Italian Ministry of Ministry of University and Research (MUR). E.B. acknowledges contribution of the Next Generation EU funds within the National Recovery and Resilience Plan (PNRR), Mission 4 - Education and Research, Component 2 - From Research to Business (M4C2), Investment Line 3.1 - Strengthening and creation of Research Infrastructures, Project IR0000034 – “STILES - Strengthening the Italian Leadership in ELT and SKA. MB acknowledges the support from the European Research Council (ERC) Advanced grant MOPPEX 833460. This project has received funding from the European Research Council (ERC) under the European Union Horizon Europe programme (grant agreement No. 101042275, project Stellar-MADE).

This paper uses the following ALMA data: ADS/JAO.ALMA\#2018.1.01205.L, ADS/JAO.ALMA\#2019.1.01566.S, ADS/JAO.ALMA\#2013.1.00291.S. ALMA is a partnership of ESO (representing its member states), NSF (USA) and NINS (Japan), together with NRC (Canada), MOST and ASIAA (Taiwan), and KASI (Republic of Korea), in cooperation with the Republic of Chile. The Joint ALMA Observatory is operated by ESO, AUI/NRAO, and NAOJ. The National Radio Astronomy Observatory is a facility of the National Science Foundation operated under cooperative agreement by Associated Universities, Inc.
\end{acknowledgements}

% WARNING
%-------------------------------------------------------------------
% Please note that we have included the references to the file aa.dem in
% order to compile it, but we ask you to:
%
% - use BibTeX with the regular commands:
%   \bibliographystyle{aa} % style aa.bst
%   \bibliography{Yourfile} % your references Yourfile.bib
%
% - join the .bib files when you upload your source files
%------------------------------------------------------------------
\bibliographystyle{bibtex/aa}{}
\bibliography{aandabib}

\onecolumn
\begin{appendix} %First appendix
\section{CS\,(5-4) channel maps}

\begin{figure*}[h]
       \centering
       \includegraphics[width=\textwidth]{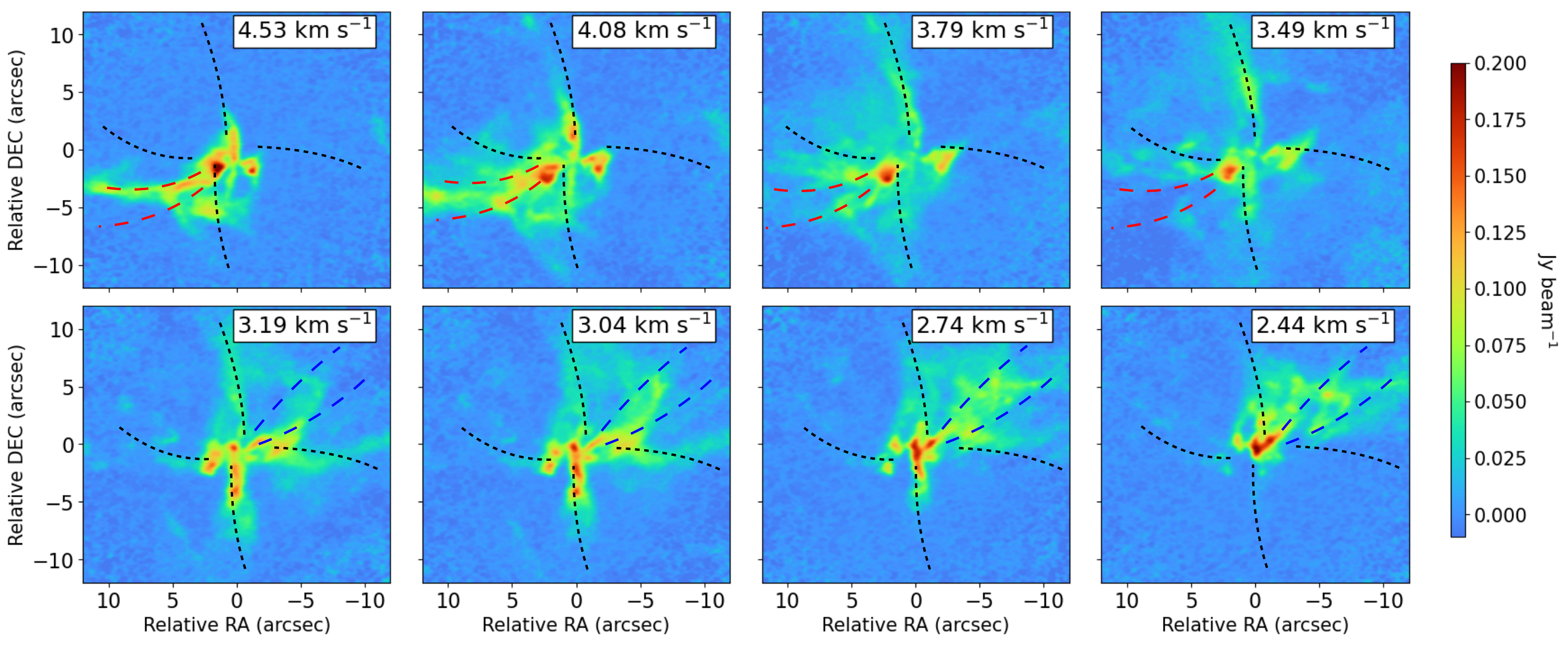}
       \caption{CS\,(5-4) channel maps: the velocity of the channel is indicated in the panel at the top right corner of each map. The dashed black line delineates the cavity of the CO outflow of \cite{alves2017}, and the red and blue dashed lines indicate the regions where the redshifted and blueshifted elongated structures are found, respectively. The synthesized beam size is of 0.49"$\times$0.42".} \label{fig:cschannelmaps}
    \end{figure*}

Figure \ref{fig:cschannelmaps} shows a selection of channel maps from the CS\,(5-4) ($E_u=35.26$\,K) line emission, which traces the CO outflow discovered by \cite{alves2017} as well as part of the elongated structures discovered by \cite{evans2023} and analyzed in this work.

% Because the optical images used in this analysis...
% \begin{figure*}%f1
% \includegraphics[width=10.9cm]{1787f23.eps}
% \caption{Shown in greyscale is a...}
% \label{cl12301}
% \end{figure*}

% In this case....
% \begin{figure*}
% \centering
% \includegraphics[width=16.4cm,clip]{1787f24.ps}
% \caption{Plotted above...}
% \label{appfig}
% \end{figure*}

% Because the optical images...

% \section{Title of Second appendix.....} %Second appendix
% These studies, however, have faced...
% \begin{table}
% \caption{Complexes characterisation.}\label{starbursts}
% \centering
% \begin{tabular}{lccc}
% \hline \hline
% Complex & $F_{60}$ & 8.6 &  No. of  \\
% ...
% \hline
% \end{tabular}
% \end{table}

% The second method produces...
\end{appendix}
\end{document}